\documentclass[showpacs,twocolumn,superscriptaddress]{revtex4}

\usepackage{doi}
\usepackage{hyperref}
\hypersetup{
	colorlinks=true,        
	linkcolor=blue,         
	citecolor=cyan,         
}

\usepackage{graphicx}
\usepackage{dcolumn}
\usepackage{bm}
\usepackage{color}

\usepackage{amsmath}
\usepackage{amssymb}

\def\a{\alpha}
\def\r{\rho}
\def\s{\sigma}
\def\t{\tau}
\def\m{\mu}
\def\n{\nu}
\def\k{\kappa}
\def\th{\theta}
\def\g{\gamma}\def\G{\Gamma}
\def\L{\Lambda}\def\l{\lambda}
\def\D{\Delta}
\def\la{\langle}
\def\ra{\rangle}
\def\o{\omega}\def\O{\Omega}
\def\d{\delta}
\def\p{\partial}

\def\half{\textstyle{\frac{1}{2}}}

\def\bdoc{\begin{document}}
	\def\edoc{\end{document}}
\def\beq{\begin{equation}}
\def\eeq{\end{equation}}
\def\bea{\begin{eqnarray}}
\def\eea{\end{eqnarray}}
\def\ben{\begin{enumerate}}
	\def\een{\end{enumerate}}
\def\la{\langle}
\def\ra{\rangle}
\def\a{\alpha}
\def\b{\beta}
\def\g{\gamma}
\def\G{\Gamma}
\def\d{\delta}
\def\D{\Delta}
\def\e{\epsilon}

\def\th{\theta}
\def\k{\kappa}
\def\l{\lambda}
\def\m{\mu}
\def\n{\nu}
\def\o{\omega}
\def\p{\pi}
\def\r{\rho}
\def\s{\sigma}
\def\t{\tau}
\def\L{{\cal L}}
\def\S{\Sigma }
\def\gsim{\; \raisebox{-.8ex}{$\stackrel{\textstyle >}{\sim}$}\;}
\def\lsim{\; \raisebox{-.8ex}{$\stackrel{\textstyle <}{\sim}$}\;}
\def\gtrsim{\gsim}
\def\lessim{\lsim}
\def\loc{{\rm local}}
\def\vm{v_{\rm max}}
\def\bh{\bar{h}}
\def\del{\partial}
\def\nab{\nabla}
\def\half{{\textstyle{\frac{1}{2}}}}
\def\fourth{{\textstyle{\frac{1}{4}}}}

\def\bD{{\bf D}}
\def\bE{{\bf E}}
\def\bF{{\bf F}}
\def\bB{{\bf B}}
\def\bP{{\bf P}}
\def\bV{{\bf v}}
\def\bv{{\bf v}}
\def\bx{{\bf x}}
\def\by{{\bf y}}
\def\bz{{\bf z}}
\def\ba{{\bf a}}
\def\bd{{\bf d}}
\def\bs{{\bf s}}
\def\bn{{\bf n}}
\def\bp{{\bf p}}

\def\O{\Omega}

\def\br{{\bf r}}
\def\bnab{{\bf \nab}}

\def\tE{\tilde{E}}
\def\tL{\tilde{L}}

\begin{document}
	
	\title{Optical Properties of an Axially Symmetric Black Hole in the Rastall Gravity}
	
	\author{Bakhtiyor Narzilloev}
	\email{nbakhtiyor18@fudan.edu.cn}
		\affiliation{Ulugh Beg Astronomical Institute, Astronomy St.  33, Tashkent 100052, Uzbekistan}
\affiliation{Akfa University, Milliy Bog' Street 264, Tashkent 111221, Uzbekistan}
	\affiliation{National University of Uzbekistan, Tashkent 100174, Uzbekistan}

	\author{Ibrar Hussain}
	\email{ibrar.hussain@seecs.nust.edu.pk}
	
	\affiliation{School of Electrical Engineering and Computer Science, National University of Sciences and Technology, H-12, Islamabad, Pakistan}
	
	\author{Ahmadjon~Abdujabbarov}
	\email{ahmadjon@astrin.uz}
	\affiliation{Shanghai Astronomical Observatory, 80 Nandan Road, Shanghai 200030, P. R. China}
	\affiliation{Ulugh Beg Astronomical Institute, Astronomy St.  33, Tashkent 100052, Uzbekistan}
	\affiliation{Institute of Nuclear Physics, Ulugbek 1, Tashkent 100214, Uzbekistan}
	\affiliation{National University of Uzbekistan, Tashkent 100174, Uzbekistan}
	\affiliation{Institute of Fundamental and Applied Research, National Research University TIIAME, Kori Niyoziy 39, Tashkent 100000, Uzbekistan}

	\author{Bobomurat Ahmedov}
	\email{ahmedov@astrin.uz}
	
	\affiliation{Ulugh Beg Astronomical Institute, Astronomy St.  33, Tashkent 100052, Uzbekistan}
	\affiliation{National University of Uzbekistan, Tashkent 100174, Uzbekistan}
	\affiliation{Institute of Fundamental and Applied Research, National Research University TIIAME, Kori Niyoziy 39, Tashkent 100000, Uzbekistan}
	

	\date{\today}
	\begin{abstract}
		This work is devoted to the study of the optical properties of the charged-rotating-NUT-Kiselev (CRNK) black hole in the Rastall theory of gravity. By investigating the motion of photons in the CRNK black hole spacetime in the Rastall gravity we show that the deflection angle of photons due to the gravitational lensing is mostly influenced by the NUT charge, parameter of the equation of state for the quintessence and the quintessential intensity. We observe that the effects of the rest of the spacetime parameters are negligible on the deflection angle. We present the shape of the shadow cast by the CRNK black hole in the Rastall gravity for various values of the spacetime parameters. We demonstrate that spin parameter $a$ of the black hole changes the position and size of the shadow of the black hole very little. While changing the values of the rest of the spacetime parameters can alter the shape and the size of the observed shadow of the black hole significantly. We observe that the radius of the black hole shadow and the deflection angle of photon are bigger for the CRNK black hole in the Rastall gravity, as compared with the case of the Kerr black hole. Lastly, we show how the spacetime parameters can change the observables $R_s$ and $\delta_s$, being the average radius of the shadow and the distortion parameter that measures the deviation of the shape of the shadow from a perfect circle with radius $R_s$, respectively. Our results show that in most scenarios, for bigger values of $R_s$ we get smaller values of the parameter $\delta_s$.
	\end{abstract}
	\pacs{04.70.Bw, 04.50.Kd, 04.70.-s}
	\maketitle
	
	
	\section{Introduction}
	\label{sec1}
To understand the gravitational fields of the ultra compact objects like black holes, the investigation of the geodesics of test particles and photons in black hole spacetimes is of interest from the astrophysical point of view. The photon motion around compact objects has a direct connection with one of the predictions of the Einstein theory of General Relativity (GR) about the light bending. According to GR, due to the curvature produced by a compact object light bends and due to this bending of light or photon beam by a gravitating source, it fabricates an effect very similar to that of a lens, called Gravitational Lensing (GL) \cite{1a, 2a}. The phenomena of GL was first observed in the solar eclipse of 1919 by Arthur Eddington. After the discovery of quasars in 1963, which are very bright and distinct objects, the GL effects are studied with deeper interest \cite{3a}. The first ever GL effect of a quasar was observed in 1979 \cite{4a}. The GL which distorts the image of some background galaxy in its line of sight, can also amplify its light and can provide images of fainter and more distant galaxies with the help of telescopes. Thus GL is a useful tool to study the nature of galaxies too far away from us to be seen with the use of technology. The GL may be also helpful in detecting the mysterious dark matter in the Universe (see for example \cite{4b}).

Using the GL as an important tool various properties of static and rotating black holes have been studied in the GR and other modified theories of gravity \cite{5a,6a,7a,8a,9a}. The impact of different fields including magnetic field and quintessential field on the GL have been thoroughly investigated in the literature \cite{10a, 11a,12a,13a,14a}. An analytic method for the GL in the spacetime field of the Schwarzschild black hole was first studied by Boazza et. al \cite{15a}. Their technique was later generalized for all spherically symmetric and statics spacetimes \cite{16a}. The effects of the spin of a black hole on the GL have been studied in a separate work by Boazza \cite{17a}. A comparative study of the GL by the Schwarzschild black hole and naked singularity has been presented by Virbhdra and Keeton, where they have observed a qualitative difference in the lensing characteristics in the two cases \cite{18a}. An approach for the study of the GL in the Kerr spacetime has been provided by Beckwith and Done \cite{19a}. The effects of charge of black holes on the GL have also been studied in rotating and non rotating spacetimes to understand the gravitational fields and the spacetime geometry in these cases \cite{20a,21a,22a}.

The Event Horizon Telescope (EHT) collaboration released the first ever image of a black hole at the centre of a nearby galaxy in 2019, which has further enhanced the interest of researches in studying the black hole features. When photon or light passes in the close vicinity of a black hole, due to the deflection by the gravitational field of the black hole the photons can move in circular orbits or rings, known as light rings (LRs). Due to the LRs the central black hole appears as a dark disc in the sky. This dark discs is known as the black hole shadow. The idea about the observation of the black hole shadow was first introduced by Falcke et al. \cite{23a}. The masses of supermassive and stellar mass black holes have been measured observationally during the last decade of the previous century \cite{24a}. Another essential parameter of a black  hole is assumed to be its spin, whose measurement is a crucial challenge in the present day astrophysics. To estimate the black hole spin, some methods have been put forwarded by different people (see for example \cite{25a,26a}). It has been proposed that the black hole shadows may be useful in estimating the spin of the background black holes \cite{23a,26a}. Black hole shadows have been investigated in different black hole spacetimes from different point of views (for example \cite{27a,28a,29a,30a}).

The generalization of the very first black hole solution of the Einstein field equations, known as the Schwarzschild black hole solution, in the presence of some physical parameters has been remain a topic of interest. After the introduction of charge and spin in the Schwarzschild black hole solution, people have further studied this solution in the presence of some other quantities including magnetic field and some dark energy candidates. In this regard Kiselev has been obtained the Schwarzschild black hole solution in the quintessence background \cite{32a}. The charged and rotating versions of the Kiselev black hole solution have also been reported in the literature \cite{32a,33a}. The rotating version of the charged Kiselev black hole solution has been further generalized in the presence of the NUT parameter and has been analysed in different context in the literature \cite{34a,35a}. In the recent years the charged rotating version of the Kiselev black hole with the NUT parameter has further been studied in the Rastall theory of gravity \cite{36a}. Recently particle dynamics have been studied in the spacetime of the charged-rotating Kiselev black hole with the NUT parameter in the Rastall theory of gravity \cite{37a}. Further the effects of the Rastall parameter and other spacetime parameters have been investigated on the fundamental frequencies of test particles at small distances from the stable circular orbits, in the charged-rotating Kiselev black hole wit the NUT parameter in the background of the Rastall theory of gravity \cite{37a}.
 {The spacetime properties of rotating black holes and exotic compact objects in the Kerr/CFT correspondence within the Rastall gravity have been studied in~\cite{36a}. In Ref.~\cite{AAA1} the Rastall gravity is generalized to the case when the conservation law of the  energy-momentum tensor is altered.  Exact cylindrical symmetric  solutions of the field equations in the Rastall theory of gravity have been obtained in~\cite{AAA2}. The properties of rapidly rotating compact stars in the Rastall gravity are studied in~\cite{AAA3}. Gravastar solution within the framework of the Rastall gravity has been explored in~\cite{AAA4} . The effects of the generalized Rastall gravity on the properties of compact objects have been analyzed in~\cite{AAA5}. It has been shown that for a suitable choice of the coupling parameter of the Rastall gravity in the presence of the quintessence parameter, the original results in the standard GR can be retrieved~\cite{AAA6}. Authors of~\cite{Sakti2019krw} have obtained a new twisted rotating black hole solution applying Demianski-Newman-Janis algorithm to the electrically and dyonically charged black hole with quintessence in the Rastall theory of gravity. Other axisymmetric solutions of the field equations of the Rastall gravity have been obtained in Refs.~\cite{AAA8, AAA9, AAA10, AAA11}. In a recent work shadow cast by a black hole surrounded by an anisotropic fluid in the Rastall theory of gravity has been studied \cite{R0}.}
Interested readers may see our previous works devoted to the investigation of spacetime properties around several massive compact objects \cite{Narzilloev19,Narzilloev:2020b,Narzilloev20c,Narzilloev20a,Narzilloev20b, Narzilloev21, Hakimov17, Narzilloev21a, Narzilloev21b, Narzilloev21c, narzilloev21d}.  {It has been demonstrated in~\cite{VISSER201883} that at the global scale the Rastall gravity can be treated as an  essential and trivial re-arrangement of the matter sector in the standard  Einstein gravity. If one compares the  Einstein and the Rastall theories of gravity then one may observe that gravity parts of the both theories are equivalent to each other. Particularly, the non-generic case of the Rastall gravity is nothing else but the standard Einstein gravity implemented with the cosmological constant. This fact also drives the motivation to study the Rastall gravity in the sense of studying the matter sector of the theory, deeply using the spacetime properties in the close vicinity of gravitational compact objects.}

In the present work we are keen in the investigation of photon motion and related phenomena of GL and black hole shadows in the spacetime of the  charged-rotating-NUT-Kiselev (CRNK) black hole in the Rastall gravity. In particular, we are interested to look at the effects of the Rastall parameter and the other spacetime parameters on the GL and shadow cast by the CRNK black hole.

This paper is organized as follows: In the next Section we discuss the spacetime structure of the CRNK black hole in the Rastall theory of gravity. In the Section III the GL of the CRNK black hole in the Rastall gravity is studied. Section IV is devoted to the analysis of shadow of the CRNK black hole in the Rastall gravity. In the last Section we give a conclusion of our work done here.

\section{Charged-Rotating-NUT-Kiselev spacetime in the Rastall gravity}\label{sec2}

The GR is a promising theory of gravity and has already passed some observational tests including the deflection of light by gravitational sources and the generation of gravitational waves from a binary system of black holes \cite{38a}. There are still some mysteries exist, for instance the quantization of gravity on a curved background and the accelerated expansion of the Universe, which cannot be well explained in the framework of GR. Therefore, alternative approaches for a well accepted theory of gravity have been proposed \cite{39a}. One such proposal is the Rastall theory of gravity which is a modification of the GR such that the matter field is non-minimally coupled with the geometry \cite{40a}. In the Rastall theory of gravity rotating and non-rotating black hole solutions have been reported in the literature \cite{41a, 42a}. Here we mention that it was claimed by Visser \cite{VISSER201883}, that the Rastall theory is equivalent to the GR. Later Darbari et. al \cite{Darabi2017coc}, have shown that the two theories of gravity i.e the GR and the Rastall theory are not equivalent and the latter one is an alternative theory of gravity and is to be considered as an open theory as compared to the GR.        	

 {The Kerr black hole, which can be specified by its mass and spin only, and hence obey the no-hair theorem of black holes, is to be considered as a strong candidate for an astrophysical black hole. However there is no direct evidence for it and the astrophysical observations made so far could not proved that the astrophysical black hole can exactly be described by the Kerr metric. Therefore the investigation of different properties of rotating black hole spacetimes in theories of gravity other the GR is a hot area of research. Charged and as well as uncharged rotating black hole in the Rastall theory of gravity have recently been discussed in the literature  \cite{42a,R2}. The recent observation of the image of the M87* black hole by the EHT collaboration provides a tool to test the features of gravity in the strong field regime and also to test the hypothesis of the no-hair theorem of black holes.}

 {Nevertheless the shadow cast by the M87* is almost in agreement with the shadow of the Kerr black hole, however, deviations from the Kerr black hole solution of the GR those arise from the different modified theories of gravity are still in the consideration \cite{R3,R4,R5,R6}. In this connection the Rastall theory of  gravity is to be considered to have some deviation from the GR in the strong field regime. intriguingly, the quantum effects arising in the strong gravity regime \cite{R7,R8}, the non-commutative effects \cite{R9}, asymptotic safe behaviour \cite{R10,R11} and fluctuations in the metric coefficients \cite{R12} are observed to have some influence on the shadow cast by a black hole. Motivated from these observations, here we investigate the optical properties of the CRNK black hole in the Rastall gravity.} The spacetime metric of the CRNK black hole in the Rastall gravity has the following form  {and can be checked as a solution of the Rastall (modified Einstein) field equations using the Mathematica package RGTC (see \cite{Sakti2019krw})}
\begin{eqnarray}\label{metric}
ds^2&=&-\frac{\Delta}{\Sigma} [dt-\{a \sin^2\theta+2 l (1-\cos\theta)\} d\phi]^2\\\nonumber
&+&\frac{\Sigma}{\Delta} dr^2+\Sigma d\theta^2+\frac{\sin^2\theta}{\Sigma} [a dt-\{r^2+(a+l)^2\} d\phi]^2\, ,
\end{eqnarray}
where
\begin{eqnarray}
\Delta&=&r^2-2 M r +a^2+e^2+g^2-l^2-\alpha r^v, \\ v
&=&\frac{1-3 \omega}{1-3 \kappa \lambda (1+\omega)}, \\ \Sigma&=&r^2+(l+a\cos\theta)^2\ .
\end{eqnarray}
The CRNK black hole solution in the Rastall gravity contains seven parameters where $M$ is the total mass and $a$ is the specific angular momentum of the black hole. The other parameters included in the spacetime metric of the CRNK black hole in the Rastall gravity are as follows: $l$ is the NUT parameter, $\alpha$ is the quintessential intensity, $\kappa\lambda$ is the Rastall gravity parameter, $\omega$ is the parameter of equation of state of the quintessence, $e$ and $g$ correspond to the electric and magnetic charge of the black hole, respectively \cite{Sakti2019krw}. One can introduce new parameter $q^2=e^2+g^2$ that involves the contribution of both the electric and the magnetic charges of the black hole to the spacetime investigated here.

 {The CRNK black hole solution derived in the Rastall
gravity may have two or three horizons depending on the
values of the different spacetime parameters involved there \cite{Sakti2019krw}.} The event horizon is determined as the coordinate singularity of the spacetime and is a null hypersurface of constant $r$. Consequently, the horizons are determined as the roots of the following algebraic equation
\begin{eqnarray}
r^2-2Mr+a^2+q^2-l^2-\alpha r^v=0\ .
\end{eqnarray}
For selected values of the parameters $\omega$ and $\kappa\lambda$, the horizons coincide to have only inner and outer parts or there is no cosmological horizon. In this specific case of the horizon, the thermodynamic properties of the black hole are easier to be explored. It is the reason why in a recent study \cite{Xu2019}, only two roots for the horizon equation are investigated. In the case when the horizon equation has more than two roots, it becomes very difficult to perform calculations, for example, to study the entropy product.
In the Table~\ref{tab1} those values of the parameters $\omega$ and $\kappa\lambda$ are presented which give two analytic roots of the horizon equation and are collected in~\cite{Sakti2019krw}.
\begin{table}[h]\label{tab1}
{	\caption{\label{tab1}{Selected values of black hole parameters $\omega$ and $\kappa\lambda$ which allow exact analytical solutions for inner and outer horizons presented in the last column.}}
	\begin{tabular}{lll}
		\hline\hline
		$\omega$, $\kappa\lambda$ &  & Horizon ($r_\pm$) \\\hline\\
		0,0 & & $\left(M+\frac{\alpha}{2}\right) \pm\sqrt{\left(M+\frac{\alpha}{2}\right)^2+l^2-a^2-q^2}$\\\\
		-1/3,0 & & $\frac{M}{1-\alpha} \pm \frac{\sqrt{M^2-(a^2+q^2-l^2) (1-\alpha)}}{1-\alpha}$\\\\
		0,1/6 & & $\frac{M}{1-\alpha} \pm \frac{\sqrt{M^2-(a^2+q^2-l^2) (1-\alpha)}}{1-\alpha}$\\\\
		-1/3,-1/2 & & $\left(M+\frac{\alpha}{2}\right) \pm\sqrt{\left(M+\frac{\alpha}{2}\right)^2+l^2-a^2-q^2}$\\\\
		1/3,0 & & $M \pm\sqrt{M^2+l^2+\alpha-a^2-q^2}$\\\hline\hline
	\end{tabular}}
\end{table}


\section{Photon geodesics and gravitational lensing}
In this section we discuss motion of photon and the phenomenon of the GL in the spacetime of CRNK black hole in the Rastall theory of gravity.

\subsection{Photon motion}

Photon geodesics can be explored using the following Hamilton-Jacobi equation of motion
\begin{eqnarray}
\frac{\partial S}{\partial \lambda}=- \frac{1}{2} g^{\alpha \beta}  \frac{\partial S}{\partial x^\alpha} \frac{\partial S}{\partial x^\beta} ,
\end{eqnarray}
where $\lambda$ defines the affine parameter. When the spacetime is axially symmetric one can write the action in the following form
\begin{eqnarray}
S=-\mathcal{E} t+ \mathcal{L} \phi+S(r,\theta)\ ,
\end{eqnarray}
were $\mathcal{E}$ and $\mathcal{L}$ are the conserved energy and the angular momentum of photon, respectively, in the axial symmetric spacetime of the CRNK black hole in the Rastall gravity. It can be easily checked that the Hamilton-Jacobi equation of motion is separable in the CRNK black hole spacetime and the components of the four velocity of a photon can be written as
\begin{widetext}
\begin{eqnarray}\label{geodesics}\nonumber
\dot{t}&=&\frac{\left((a+l)^2+r^2\right) \left(\mathcal{E} \left((a+l)^2+r^2\right)-a
   \mathcal{L}\right)+\Delta  \chi  \csc ^2\theta  (\mathcal{L}-\chi
   \mathcal{E})}{\Delta  \Sigma },\\
\dot{\phi}&=&\frac{a \left(\mathcal{E} \left((a+l)^2+r^2\right)-a
   \mathcal{L}\right)+\Delta  \csc ^2\theta  (\mathcal{L}-\chi
   \mathcal{E})}{\Delta  \Sigma },\\\nonumber
f(r)=\dot{r}&=&\frac{1}{\Sigma } \sqrt{\left(\mathcal{E} \left((a+l)^2+r^2\right)-a \mathcal{L}\right)^2-\Delta
   \mathcal{K}},\\\nonumber
\dot{\theta}&=&\frac{1}{\Sigma } \sqrt{\mathcal{K}-\frac{ (a
   \mathcal{E} \cos 2 \theta +4 l \mathcal{E} \cos \theta +2 \mathcal{L}-\mathcal{E} (a+4 l))^2}{8 (1+\cos
   \theta ) \sin ^2\frac{\theta }{2}}}\ ,
\end{eqnarray}
\end{widetext}
where $\mathcal{K}$ is the Carter constant, and notation $\chi=a \sin ^2\theta +2 l (1-\cos \theta )$ is introduced for convenience.


\subsection{Gravitational lensing}

\begin{figure*}[t!]
	\begin{center}
		\includegraphics[width=0.9\linewidth]{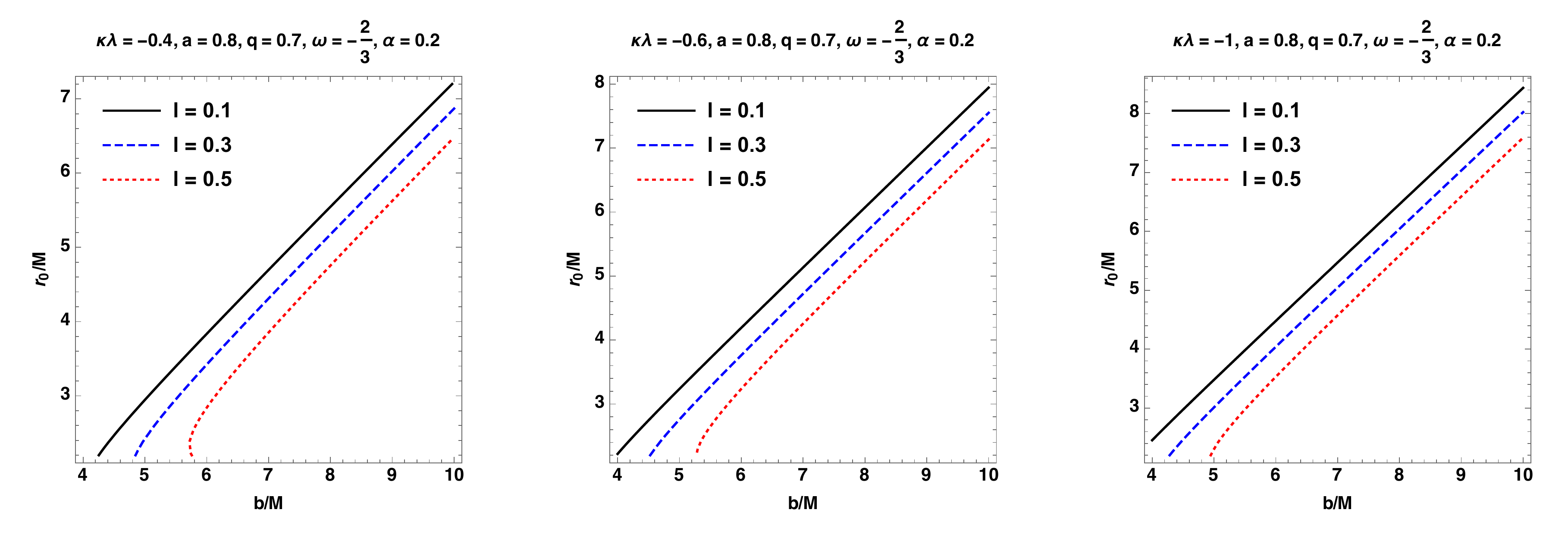}
		
		\includegraphics[width=0.87\linewidth]{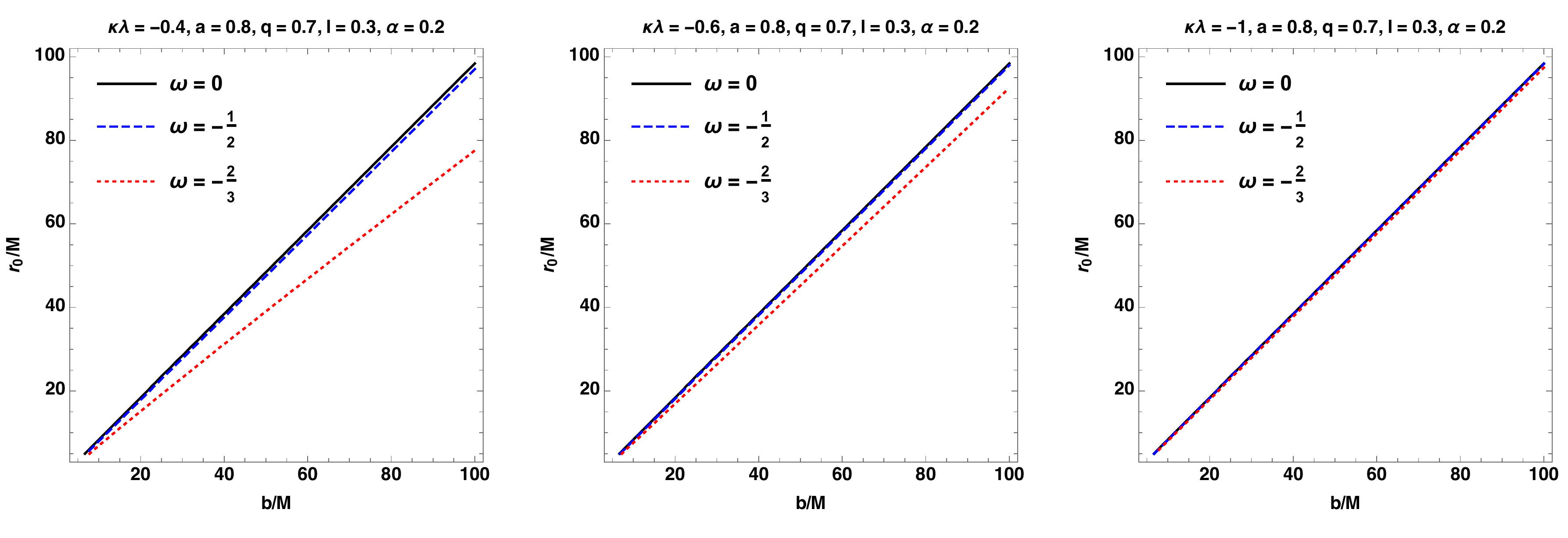}
		
		\includegraphics[width=0.9\linewidth]{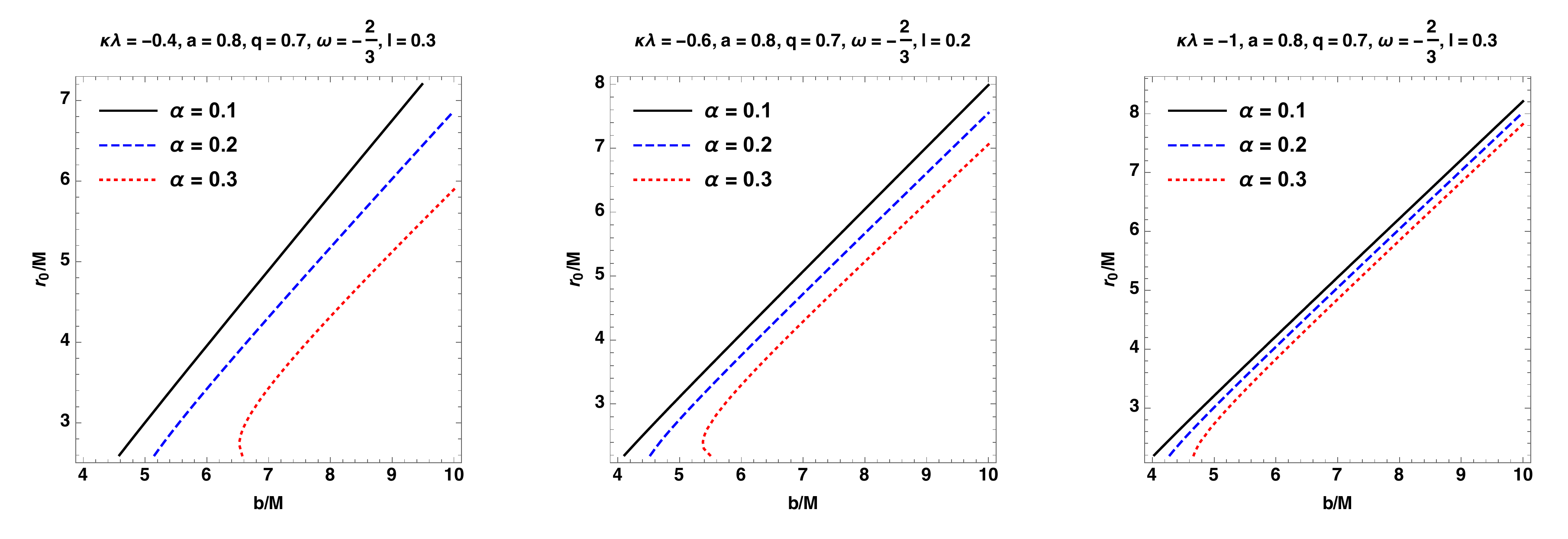}
		
		\includegraphics[width=0.9\linewidth]{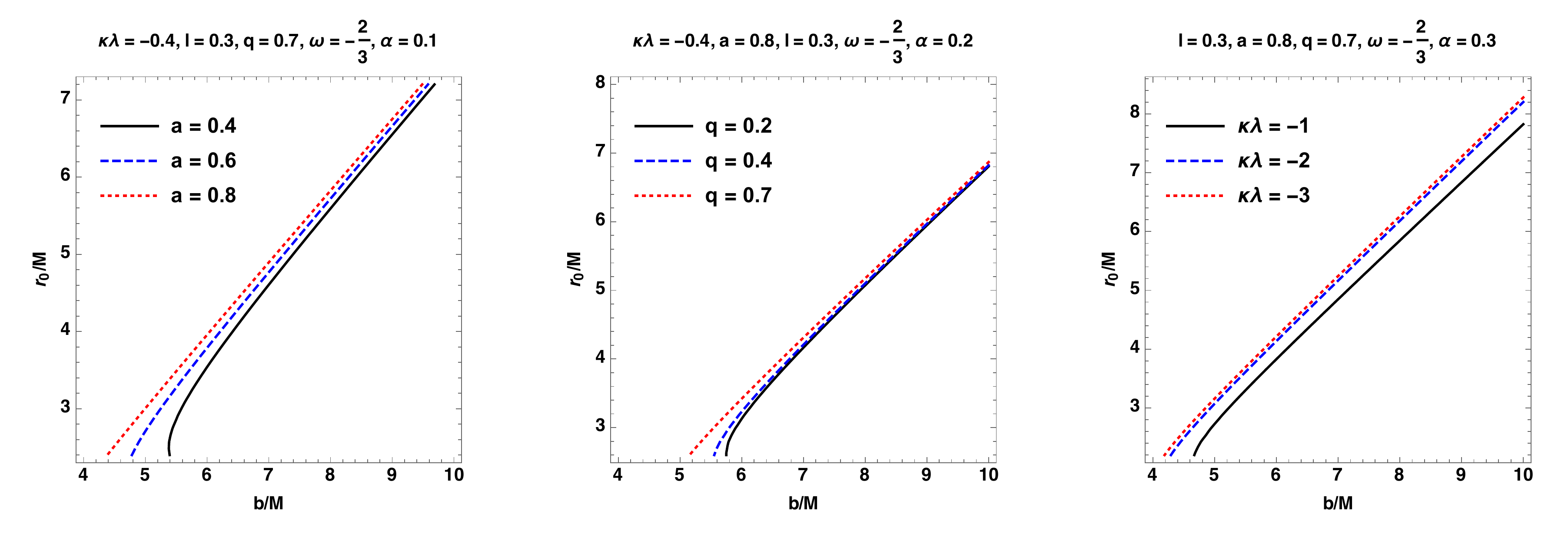}
	\end{center}
	\caption{The dependence between the radius of closest approach $r_0$ and the impact parameter $b$ for the different values of the black hole parameters. \label{r0_b}}
\end{figure*}

In this subsection we explore the lensing of photons caused by the gravitational field of the central CRNK black hole in the Rastall gravity. Assume that a photon is coming from infinity in the equatorial plane of the CRNK black hole in the Rastall gravity with impact parameter defined as $b=\mathcal{L}/\mathcal{E}$ and going back to asymptotic infinity. Using the condition $\dot{r}=0$, one can get the relationship between the radius of closest approach $r_0$ and the impact parameter from the third equation given in the system \eqref{geodesics}. This relation between $r_0$ and $b$ is demonstrated in Fig.\ref{r0_b} for the chosen values of the spacetime parameters. In the first row of the Fig.\ref{r0_b} the lines for the chosen values of the NUT parameter $l$ and the Rastall parameter $\kappa\lambda$ are presented while the other spacetime parameters are kept fixed. It can be seen from the behavior of the lines in the Fig.\ref{r0_b} that increase of the NUT parameter reduces the radius of the closest approach of photons for the fixed value of the impact parameter. This in turn means that the increase of the NUT parameter increases the intensity of the gravitational field of the central black hole. A decrease in the value of the parameter $\kappa\lambda$ however shows that (from the left figure to the right in the same row) the radius of the closest approach of photons goes up slightly, reducing the effect of the  gravitational field of the central black hole. In the second row of the Fig.\ref{r0_b} the change in the lines for the selected values of the parameters $\omega$ and $\kappa\lambda$ are given. It is shown that at large distances the behavior of the graph is almost linear and for the values of $\omega$ being close to $0$, the radius of the closest approach is nearly the same as the impact parameter. In the same figure it can be also seen that a decrease in the value of the parameter $\kappa\lambda$ makes the values of $\r_0$ and $b$ almost equal for smaller values of the parameter $\omega$. In the third row of the Fig.\ref{r0_b}, the plots correspond to the case of different values of $\alpha$ and $\kappa\lambda$. It is shown that an increase in the parameter $\alpha$ shifts the lines towards the smaller radius of the closest approach of photons, making the gravitational field of the central object stronger. A decrease in the value of the parameter $\kappa\lambda$ puts the lines closer to each other and shifts the them up, slightly decreasing the gravitational field of the black hole. In the first plot of the last row in the Fig.\ref{r0_b}, the behavior of the lines for the different values of the spin parameter $a$ is presented. One can see a slight change in the lines with the increase in the value of the spin parameter $a$ and the changes become significant only in the close vicinity of the CRNK black hole in the Rastall gravity. In the second plot of the last row in the Fig.\ref{r0_b}, the behavior of the lines for different values of the electromagnetic charge parameter $q$ is presented and can be explained in the same way as it is explained above for the different values of the spin parameter $a$. It means that the behavior of the plots is similar for both parameters $a$ and $q$. In the last plot of the same figure, the case of different values of the parameter $\kappa\lambda$ is demonstrated. It shows that the decrease of the parameter $\kappa\lambda$ indeed makes the radius of the closest approach $r_0$, bigger and hence decreases the gravitational attraction between the central black hole and photon.

\begin{figure*}[h!]
	\begin{center}
		\includegraphics[width=0.31\linewidth]{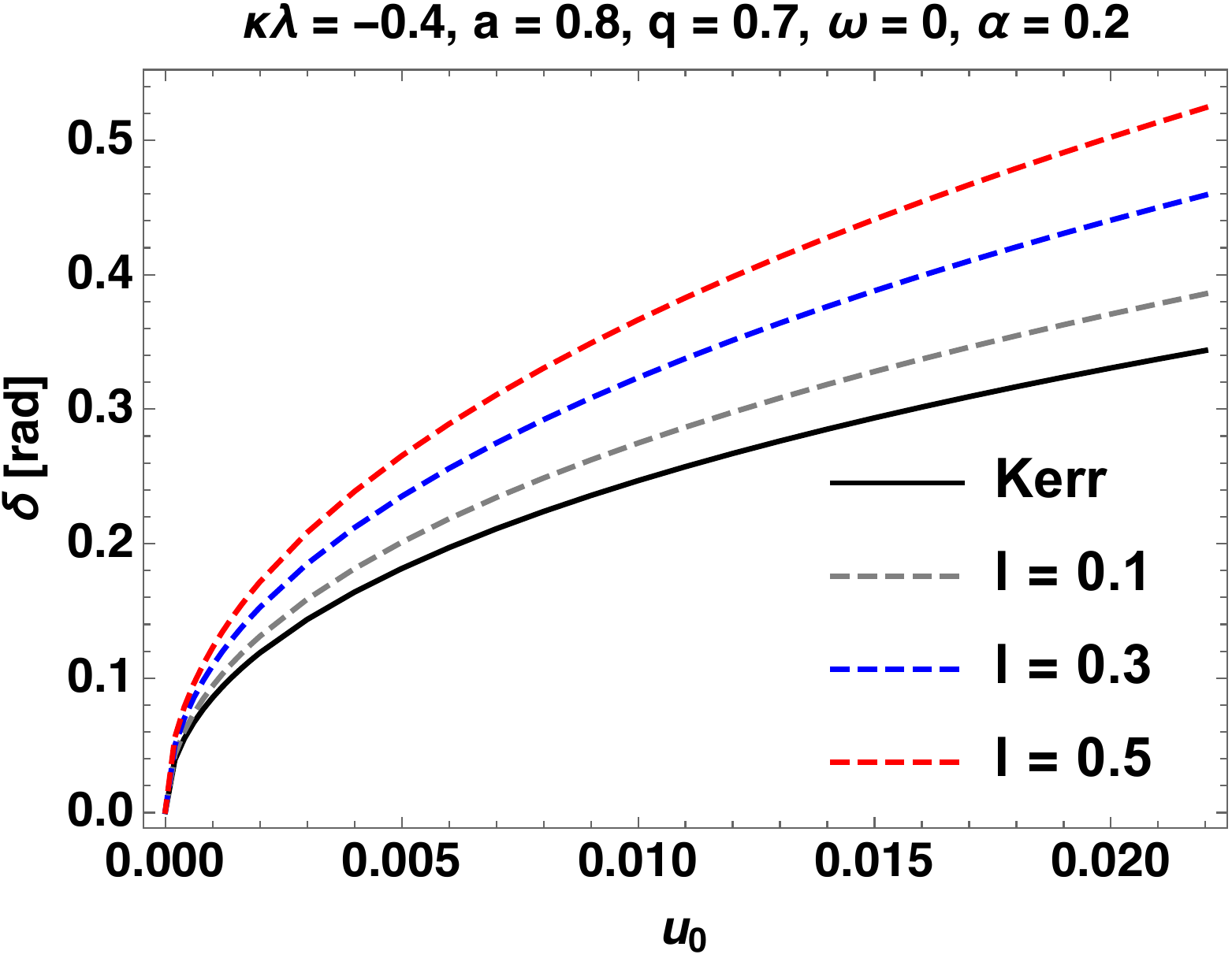}
		\includegraphics[width=0.31\linewidth]{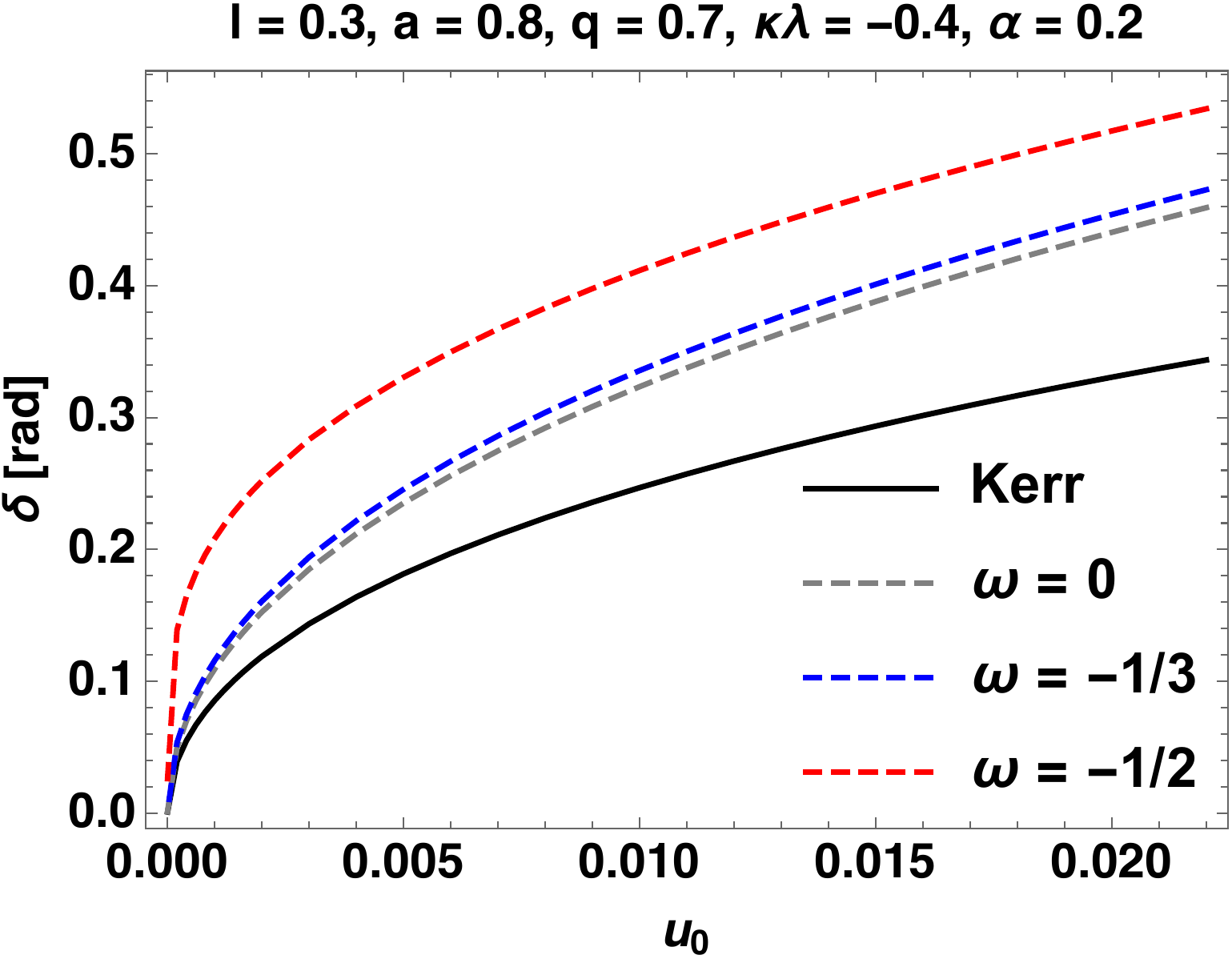}
		\includegraphics[width=0.31\linewidth]{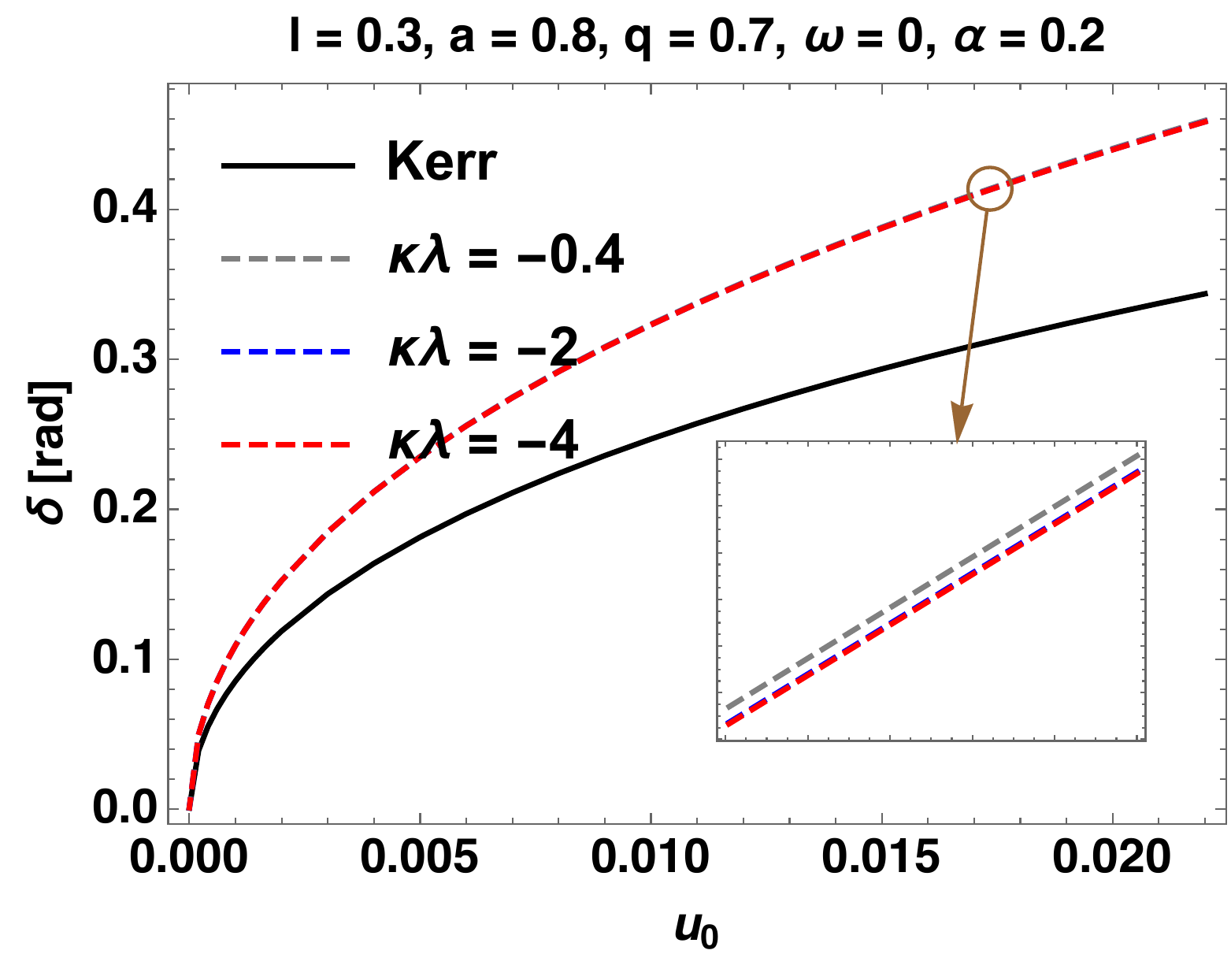}
		\includegraphics[width=0.31\linewidth]{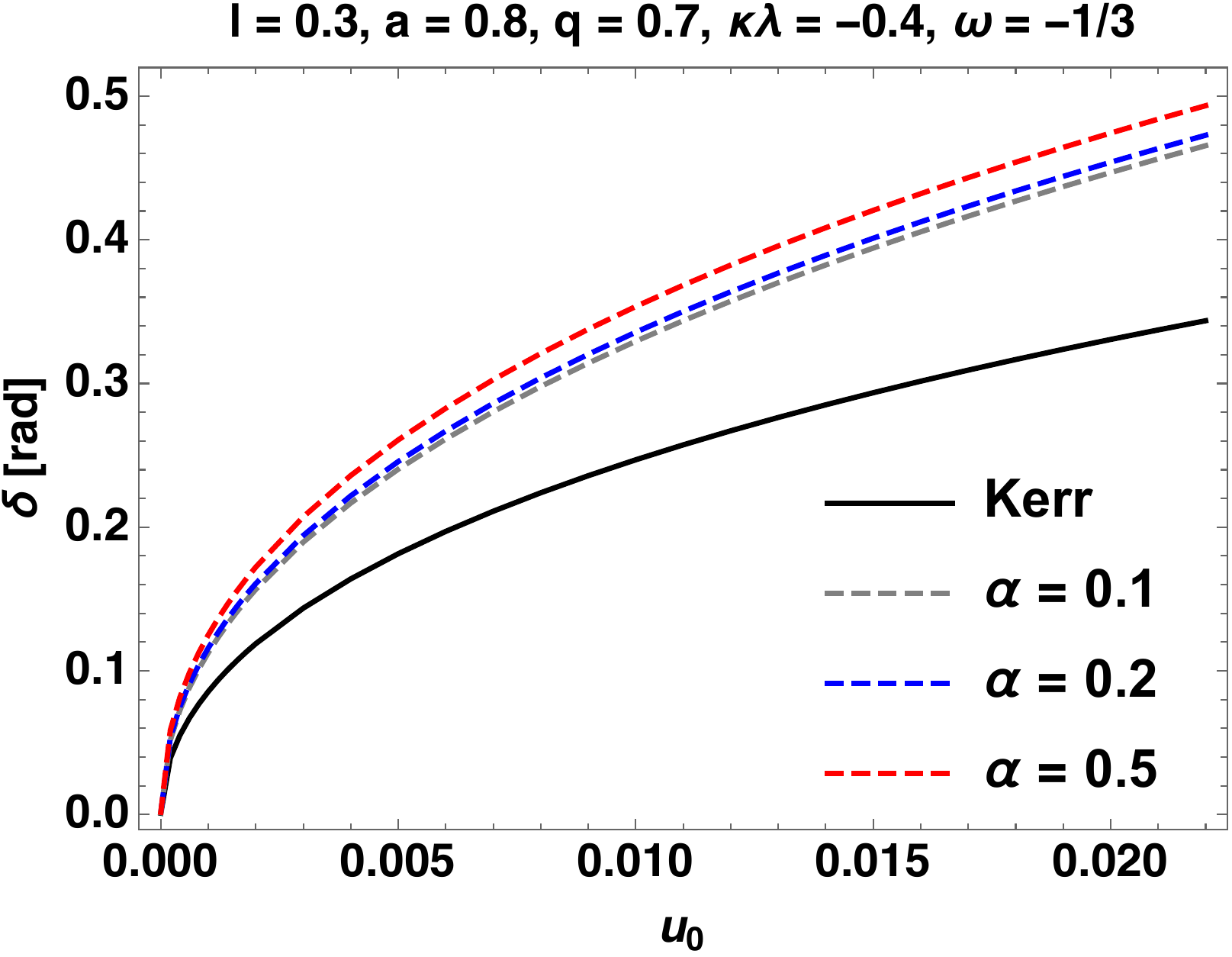}
		\includegraphics[width=0.31\linewidth]{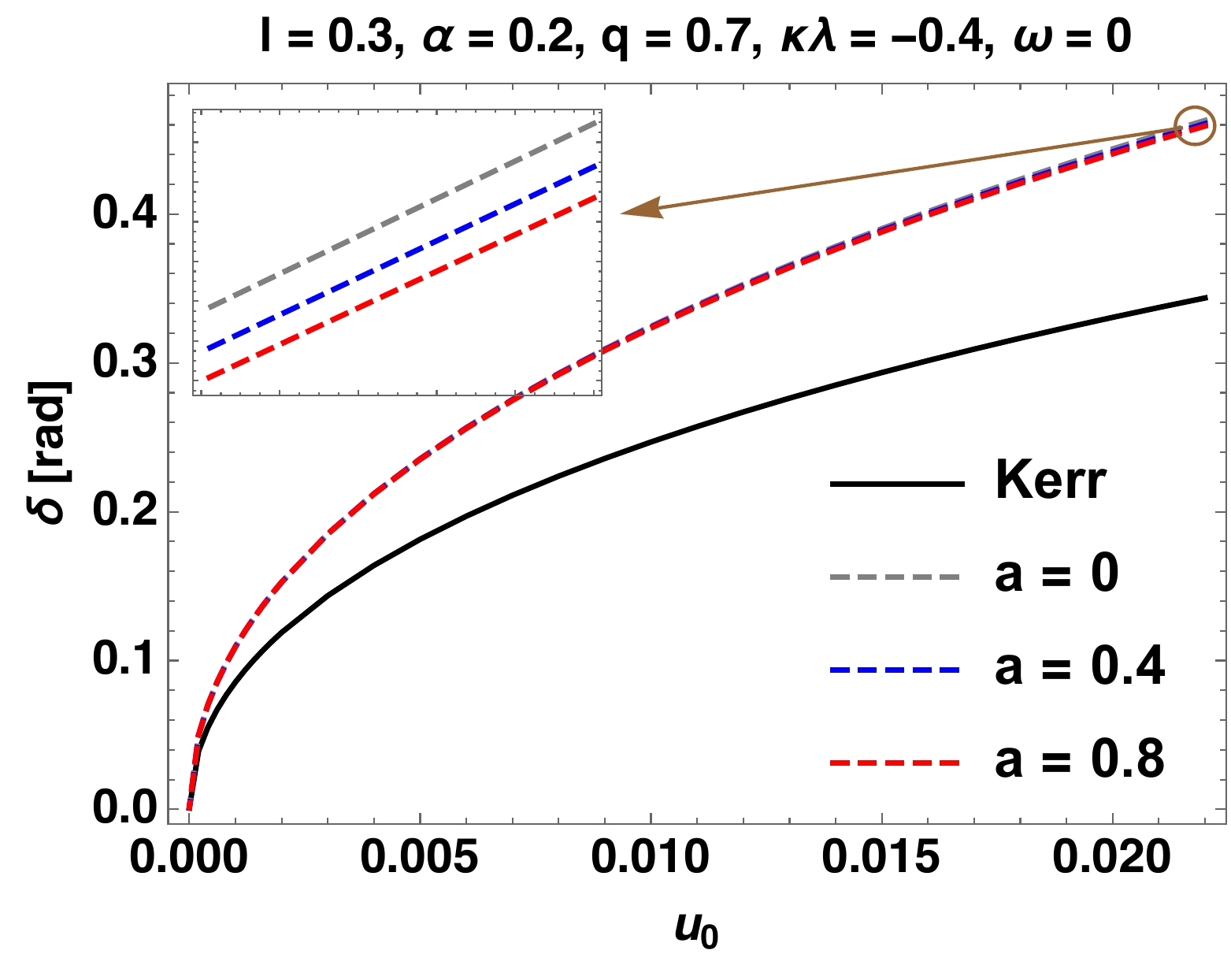}
		\includegraphics[width=0.31\linewidth]{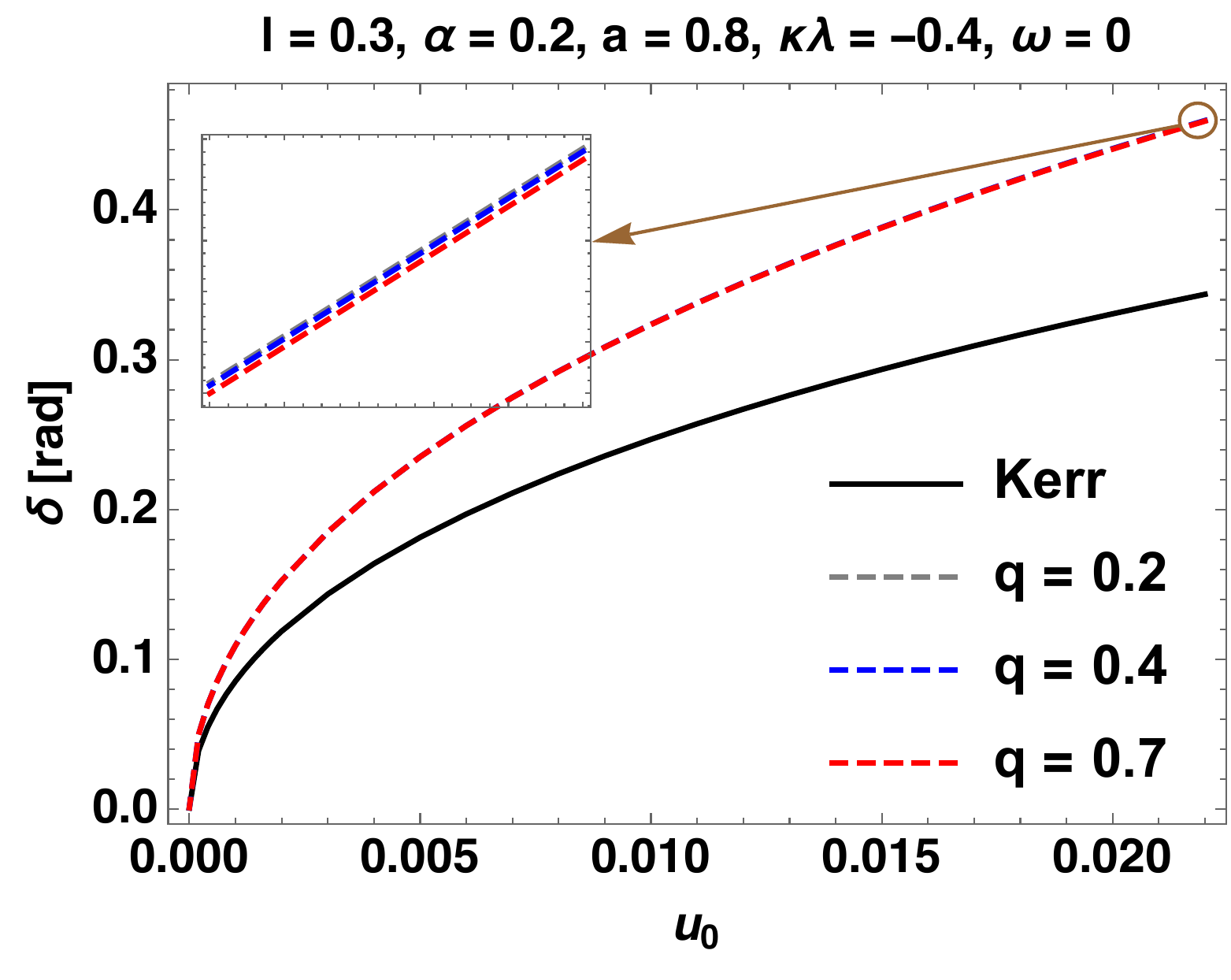}
	\end{center}
	\caption{The dependence between the bending angle of photons $\delta$ and the parameter of closest approach $u_0$. \label{d_u0}}
\end{figure*}

Using the second and third expressions in \eqref{geodesics} for the case of equatorial plane ($\theta=\pi/2$), one can get the following dependence between the radial coordinate $r$ and the angle $\phi$:
\begin{widetext}
\begin{eqnarray}
\frac{d \phi}{dr}=\frac{\Sigma \left(a \left(a^2 +a (2 l -b)+
	\left(l^2+r^2\right)\right)+\Delta  ((-a-2 l)+b)\right)}{\sqrt{\left(a^2 +a (2 l -b)+
	\left(l^2+r^2\right)\right)^2-\Delta  (a +2 l
	-b)^2}} .
\end{eqnarray}
\end{widetext}
In order to simplify the further calculations one can introduce a new variable being equal to the inverse of the radial coordinate as $u=1/r$. For this new variable the expression above can be written in the following form $$\frac{d \phi}{du}=-\frac{d \phi}{dr} \frac{1}{u^2}$$. By integrating this expression one can obtain the dependence between the bending angle of photon and the inverse of the radius of closest approach $u_0=1/r_0$ for different values of the spacetime parameters as
\begin{eqnarray}
\delta=2 \int_0^{u_0} \frac{d \phi}{du}du - \pi.
\end{eqnarray}
It is difficult to evaluate this integral analytically. However, one can evaluate the above integral numerically and obtain the desired dependence as presented in Fig.~\ref{d_u0}. Notice that since $u_0$ is the inverse of the radius of the closest approach, bigger values of this parameter corresponds to the photon that approaches  closer to the central black hole. One can see from the first plot in the first row of the Fig.~\ref{d_u0}, that bigger values of the NUT parameter cause the bending angle to get bigger values, that in turn corresponds to the stronger gravitational field of the CRNK black hole in the Rastall gravity. The second plot of the same row in the same figure shows the dependence of the bending angle of photons for the change in the parameter $\omega$. It can be seen that smaller values of this parameter make the gravitational field stronger i.e. the trajectory of photon is going to be more twisted in the spacetime of the CRNK black hole in the Rastall gravity. The third plot in the first row of the same figure demonstrates how the change in the parameter $\kappa\lambda$ modifies the bending angle of photons propagating in the spacetime of the CRNK black hole in the Rastall gravity. One can observe  a very slight decrease in the values of the bending angle with the decrease of the parameter $\kappa\lambda$. The change of the bending angle with the increase of the parameter $\alpha$ is shown in the first plot of the second row of the Fig.~\ref{d_u0}. One can see that increase of this parameter bends the light stronger which means that the gravitational field becomes stronger for bigger values of the parameter $\alpha$. In the second and third pictures of the second row of the Fig.~\ref{d_u0}, one can see a tiny decrease of the bending angle of photons with the rise of the spin and electromagnetic charge parameters $a$ and $q$, respectively. In all plots the obtained results are compared with the case of the Kerr spacetime and it is seen that the CRNK black hole in the Rastall gravity deflects the light considerably stronger than the Kerr black hole with the same mass parameter $M$.

\begin{figure*}[t!]
	\begin{center}
		\includegraphics[width=0.91\linewidth]{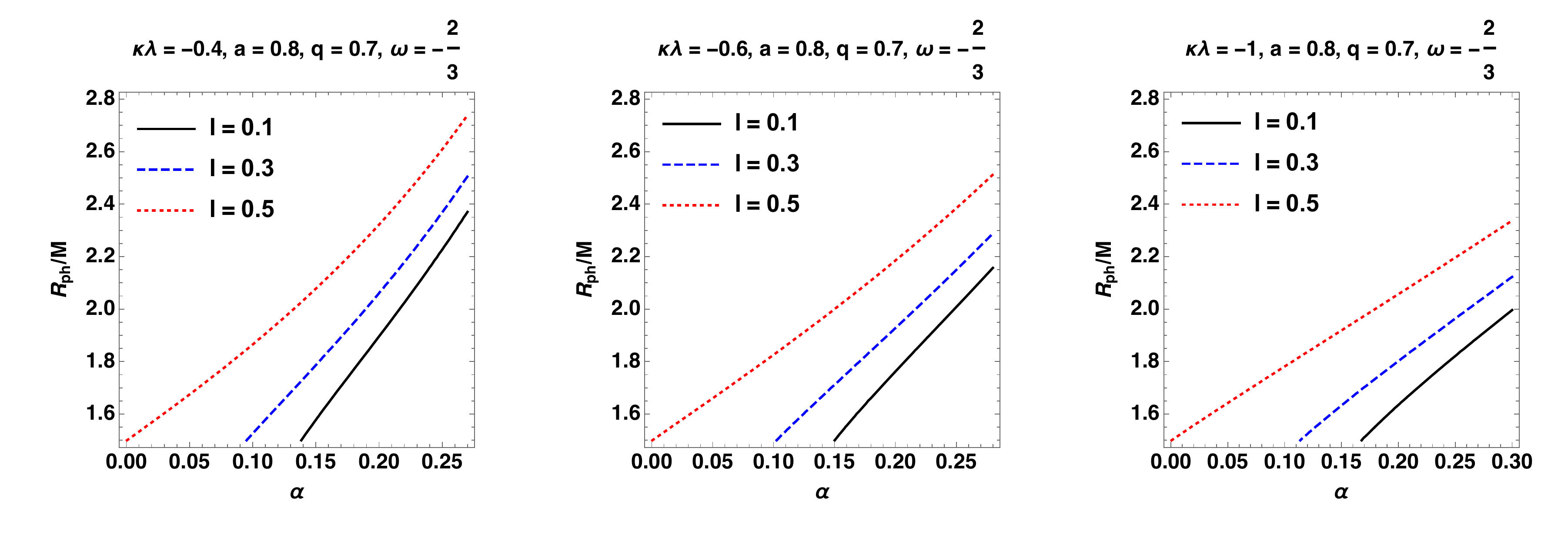}
		
		\includegraphics[width=0.87\linewidth]{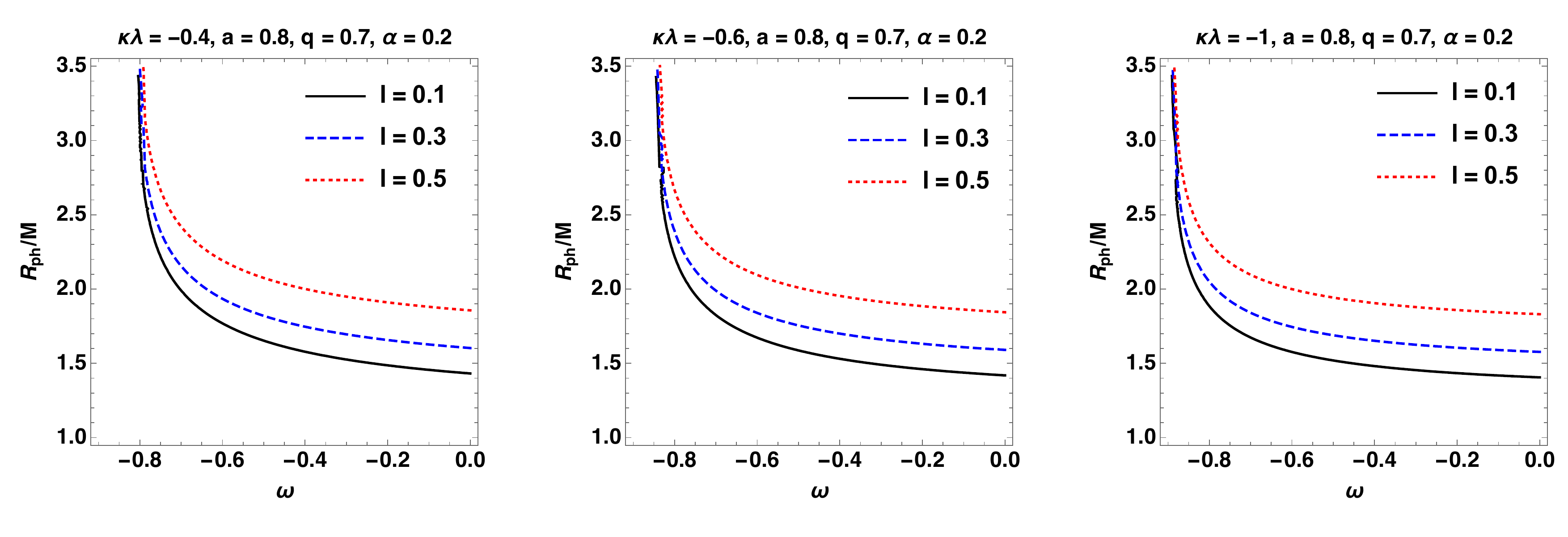}
		
		\includegraphics[width=0.9\linewidth]{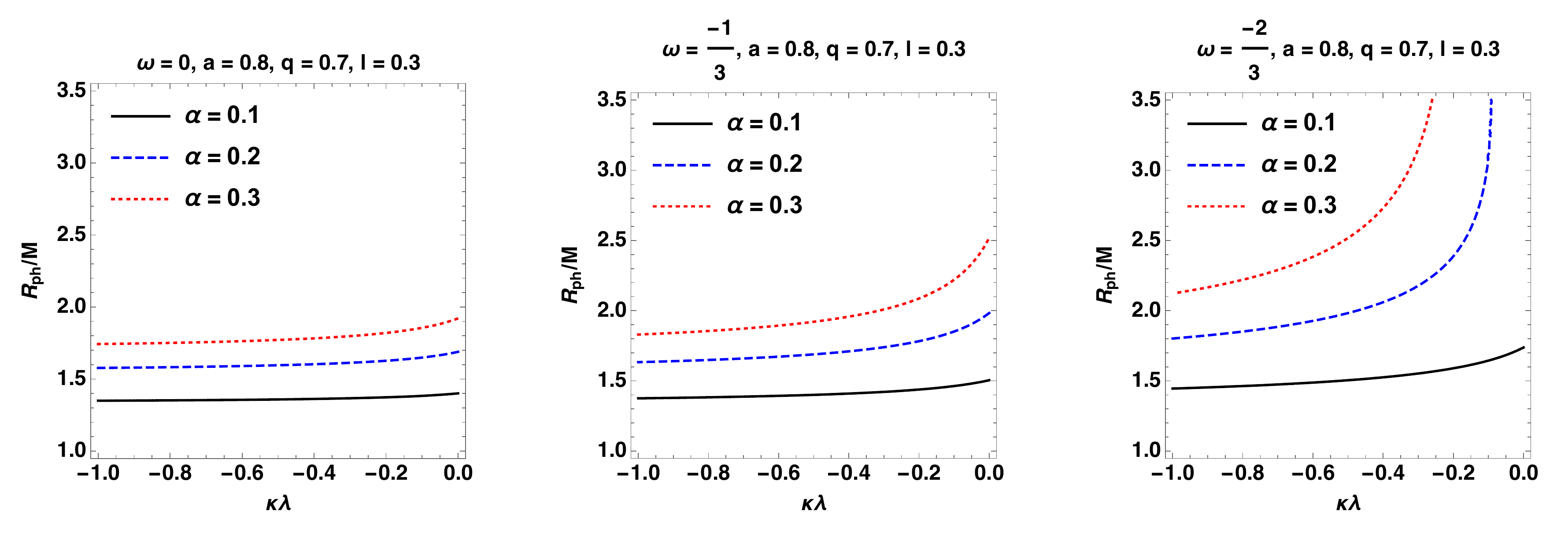}
		
		\includegraphics[width=0.9\linewidth]{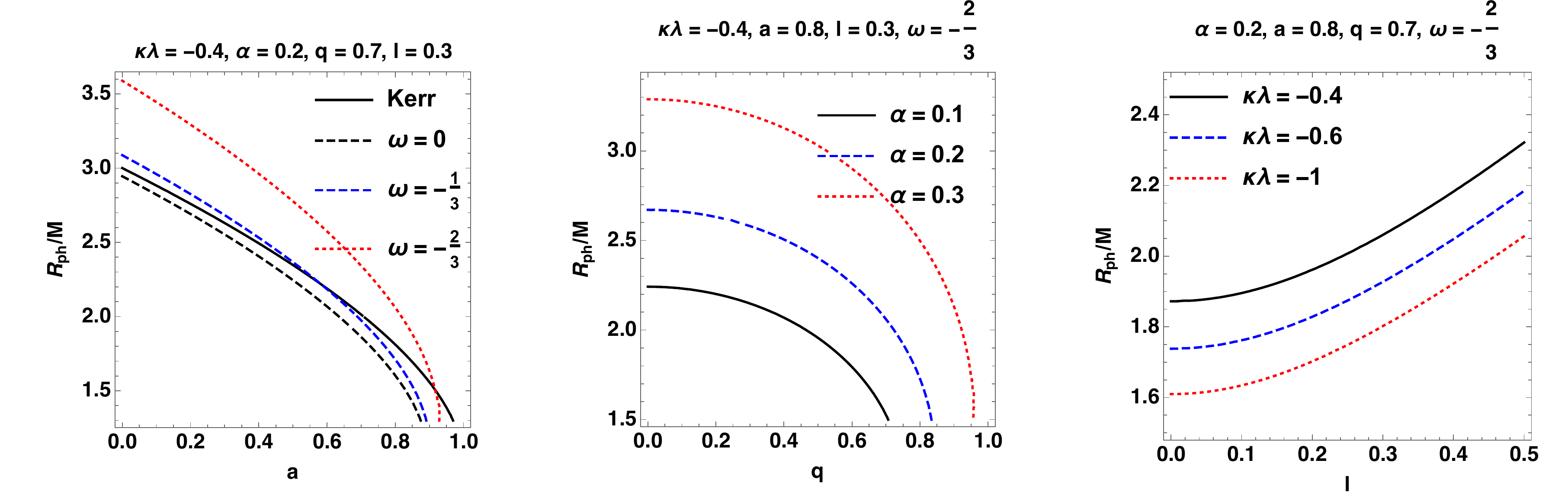}
	\end{center}
	\caption{Dependence of the photon sphere radius on the spacetime parameters of the CRNK black hole in the Rastall gravity. \label{phS}}
\end{figure*}

Before starting the investigation of the shadow formed around the CRNK black hole in the Rastall gravity let us first explore the dependence of the photon sphere on the spacetime parameters. This can be achieved by solving the following conditions
\begin{eqnarray}
f(r)&=&0\ , \\
f'(r)&=&0\ ,
\end{eqnarray}
where the form of the function $f(r)$ is given in \eqref{geodesics} and prime denotes the derivative with respect to the radial coordinate  $r$. The dependence of photon sphere on the spacetime parameters is demonstrated in Fig.~\ref{phS}. In the first plot of the first figure one can see an increase in the radius of the photon sphere with the increase of $\alpha$ and the NUT parameter $l$ which is equivalent to the strengthening of the gravitational field around the CRNK black hole in the Rastall gravity. Indeed, when the gravity is stronger photons can approach closer to the central black hole forming the circular trajectories and the strengthening of the gravitational field would increase the radius of such trajectories. Comparing this plot with the second and third plots in the same row of the Fig.~\ref{phS}, one can see a decrease in the radius of the photon sphere with the decrease of the parameter $\kappa\lambda$. This decreasing behaviour ensures once again that a decrease in the parameter $\kappa\lambda$ weakens the gravitational effects around the CRNK black hole in the Rastall gravity. In the second row of the same figure (from left to right) the change in the radius of the photon sphere, with the change of the parameter $\omega$, for fixed values of the NUT parameter $l$ and for different values of the parameter $\kappa\lambda$ can be observed. It is clearly demonstrated that decrease of the parameter $\omega$ makes the radius of the photon sphere to grow up exponentially when $\omega$ approaches $\omega\simeq-0.8$. In the third row of the same figure (from left to right) the dependence of the radius of the photon sphere on the parameter $\kappa\lambda$ is shown for some fixed values of the parameter $\alpha$ and for the different values of the parameter $\omega$. One can see that when $\omega$ is close to 0, then there is a small change in the radius of the photon sphere. It is also evident that if the values of the parameter $\omega$ are decreasing then the change in the radius of the photon sphere with the change of the parameter $\kappa\lambda$ becomes considerable only for bigger values of the parameter $\alpha$. In the first plot of the last row in the Fig.~\ref{phS}, the decrease in the radius of the photon sphere with the increase in the values of the spin parameter $a$ is plotted and then it is compared with the Kerr spacetime for different values of of the parameter $\omega$. One can see similar behavior of the lines when $\omega$ is not far from $0$. It is observed that when $\omega$ gets small enough, the difference from the Kerr case becomes apparent. In the second plot of the same row in the Fig.~\ref{phS} the dependence of the radius of the photon sphere on the electromagnetic charge parameter $q$ is presented for the selected values of the parameter $\alpha$. It can be seen that increase of the electromagnetic charge $q$ reduces the radius of the photon sphere. In the last figure the increase in the size of the photon sphere with the rise of the NUT parameter $l$ is shown for the various values of the parameter $\kappa\lambda$. The decrease of the radius of the photon sphere with the decrease of the parameter $\kappa\lambda$ is clearly observed in this picture. Consequently, the Rastall parameter $\kappa\lambda$ amplifies the strength of the gravitational field of the CRKN black hole in the Rastall gravity.

In this section we observed that the increase of the black hole parameters $l, \alpha,$ and decrease of the parameter $\omega$ make the interaction between the central black hole and photon stronger. On the contrary, the increase of the parameters $a$, $q$ and decrease of the parameter $\kappa\lambda$ makes it weaker. Now we plan to  investigate how these spacetime parameters affect the shadow around the CRNK black hole in the Rastall gravity.

\section{Black hole shadow}

\begin{figure}[h]
    \begin{center}
    \includegraphics[width=0.96\linewidth]{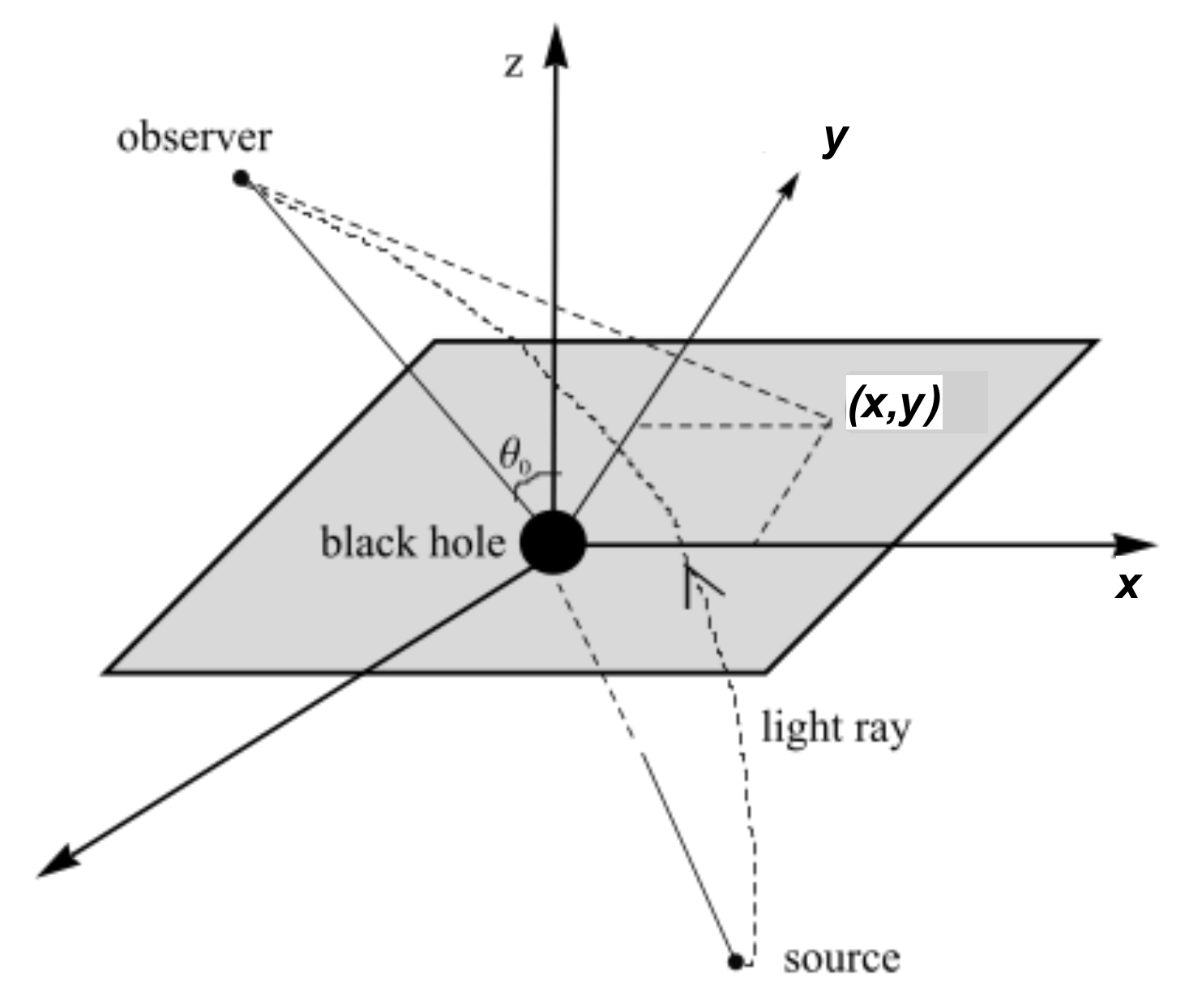}
    \end{center}
    \caption{The gravitational lens object between the source and an observer. A reference coordinate system with the black hole at the origin is set up by a distant observer. The Boyer-Lindquist coordinates tend to the Cartesian coordinate system at infinity. The black hole is rotating around the $z$ axis from the selected reference frame as seen by a distant observer at infinity. In the chosen system, the line joining the origin with the observer is orthogonal to the $xy$-plane. The vector being tangent to an incoming light ray defines a straight line intersecting the $xy$-plane at the point ($x$, $y$).}
    \label{pic1}
\end{figure}

\begin{figure}[h]
    \begin{center}
    \includegraphics[width=0.96\linewidth]{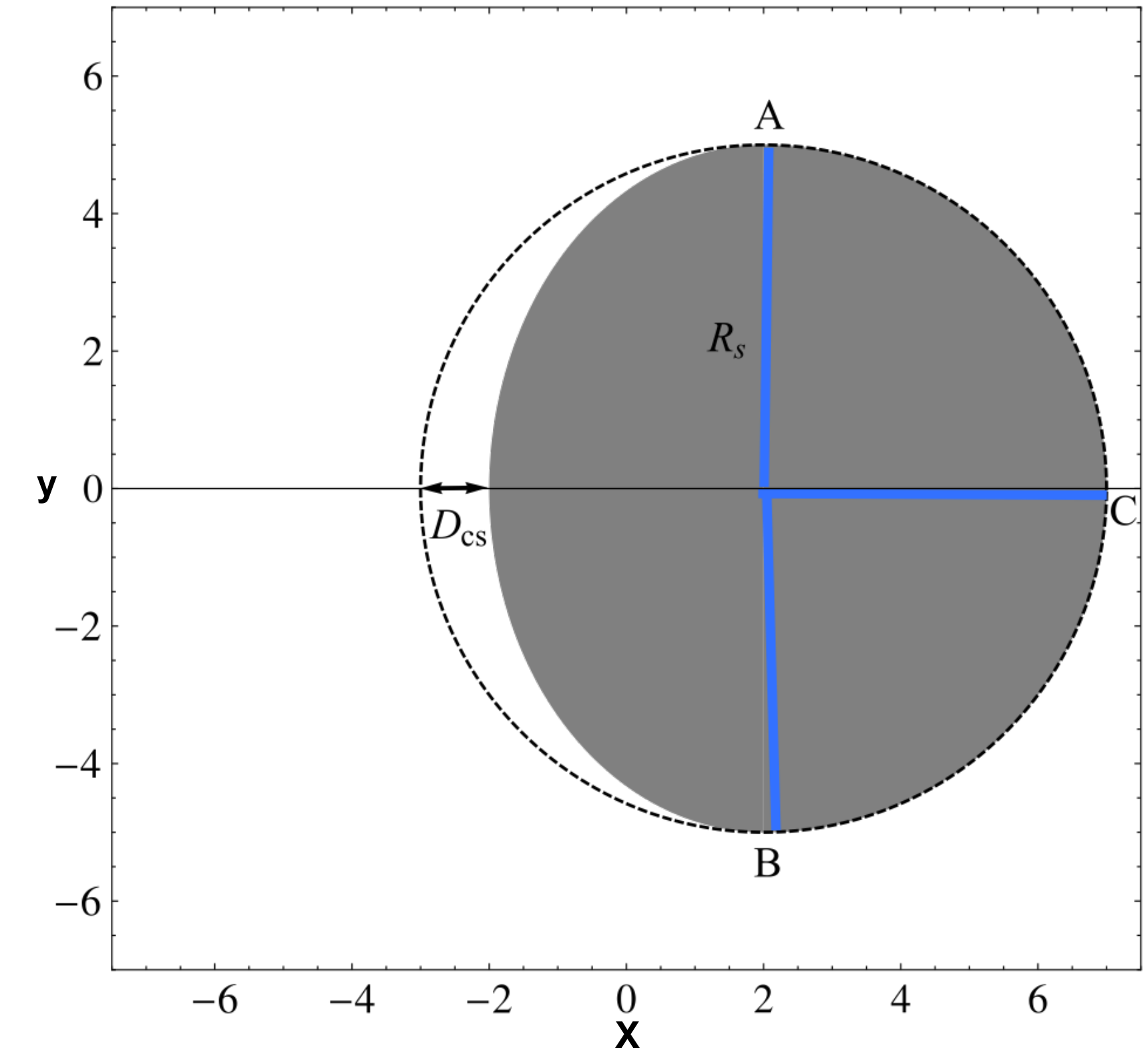}
    \end{center}
    \caption{The apparent shape of a black hole is described by two observables as the shadow radius $R_s$ and the distortion parameter $\delta_s$, approximated by a distorted circle. The parameter $D_{cs}$ is equal to the difference between the shadow and the left endpoints of the circle \cite{R13}.}
    \label{pic}
\end{figure}

In order to describe the shadow of the black hole one may use the celestial coordinates defined as (see for example \cite{Vazquez2003zm}):
\begin{eqnarray}
x=\lim_{r_0\rightarrow\infty} \left(-r_0^2 \sin\theta_0 \frac{d\phi}{dr}\right),
\end{eqnarray}
and
\begin{eqnarray}
y=\lim_{r_0\rightarrow\infty} r_0^2 \frac{d\theta}{dr},
\end{eqnarray}
where  $r_0$ is the distance between an observer and a black hole and $\theta_0$ is the angular position of the axis of rotation of the black hole with respect to the line of sight of the observer. The schematic representation of the celestial coordinates is shown in~Fig.~\ref{pic1}. In Fig.~\ref{pic1} the coordinates $x$ and $y$ describes the position of the points of the image with respect to the axis of symmetry and equatorial plane, respectively. For the spacetime~\eqref{metric}, using $d\phi/dr$ and $d\theta/dr$ given in (\eqref{geodesics}), one may obtain the following expressions for the coordinates $x$ and $y$ as functions of the impact parameter $b$ and parameter $l$ as
\begin{eqnarray}\label{eqx}
x&=&-(b_1-2l+2l\cos\theta_0) \csc\theta_0,
\\
y&=&\pm\sqrt{b_2-\frac{ (a \cos 2 \theta -a+2
   b_1+4 l \cos \theta-4 l)^2}{8 (\cos \theta+1) \sin ^2\frac{\theta }{2}}}.\ \ \label{eqy}
\end{eqnarray}
In Eqs.~(\ref{eqx}) and (\ref{eqy}) the parameters $b_1=\mathcal{L}/\mathcal{E}$ and $b_2=\mathcal{K}/\mathcal{E}^2$ are the impact parameters for general orbits around the black hole (we refer the readers to \cite{Vazquez2003zm} for the detailed calculations of these parameters for the Kerr spacetime). These parameters can be found from the conditions $\dot{r}=0=\partial_r \dot{r}$ and have the following form
\begin{eqnarray}
b_1&=&\frac{1}{2 a r (M-r)+a \alpha  v r^v} (2 r (M (a+l)^2\\\nonumber
&+&r (a-3 l) (a+l)-3 M r^2+2 q^2 r+r^3)\\\nonumber
&+&\alpha  r^v \left(v
   (a+l)^2+r^2 (v-4)\right)),\\\nonumber\\
   b_2&=&\frac{16 r^4 \left(a^2-l^2-2 M r+q^2-\alpha  r^v+r^2\right)}{\left(2 r (M-r)+\alpha  v
   r^v\right)^2}.
\end{eqnarray}
Now one may introduce two observables in order to characterize the shape of the shadow of a rotating black hole. We choose three points on the line describing the border of the shadow, namely (A), (B), and (C) corresponding to top, bottom, and rightmost positions of the image presented in the Fig.~\ref{pic1}. Note that the point (C) corresponds to the unstable retrograde equatorial circular orbits observed by a distant observer. The radius of the reference circle passing through these three points can be defined as the radius of the black hole shadow $R_s$. Using the dent in the left hand side of the shadow one may define the deflection $D_{cs}$ representing the difference between the left endpoints of the reference circle and of the image of the black hole shadow~(see the Fig.~\ref{pic}). Using the above definitions one may introduce the dimensionless distortion parameter as $\delta_s=D_{cs}/R_{s}$. We may now adopt $R_s$ and $\delta_s$ as two observables describing the shadow of rotating black holes~\cite{Hioki09}. If the distant observer located on the equatorial plane then the inclination angle is $\theta_0=\pi/2$. Consequently, one may get the maximum gravitational effects on the shadow of the black hole due to the rotation. The inclination angle corresponding to supermassive black hole at the center of our Galaxy is also expected to be close to $\pi/2$. In this particular case we have
\begin{eqnarray}
x&=&-b_1+2l,
\\
y&=&\pm\sqrt{b_2-(a-b_1+2l)^2}.
\end{eqnarray}

\begin{figure*}[h!]
	\begin{center}
		\includegraphics[width=0.36\linewidth]{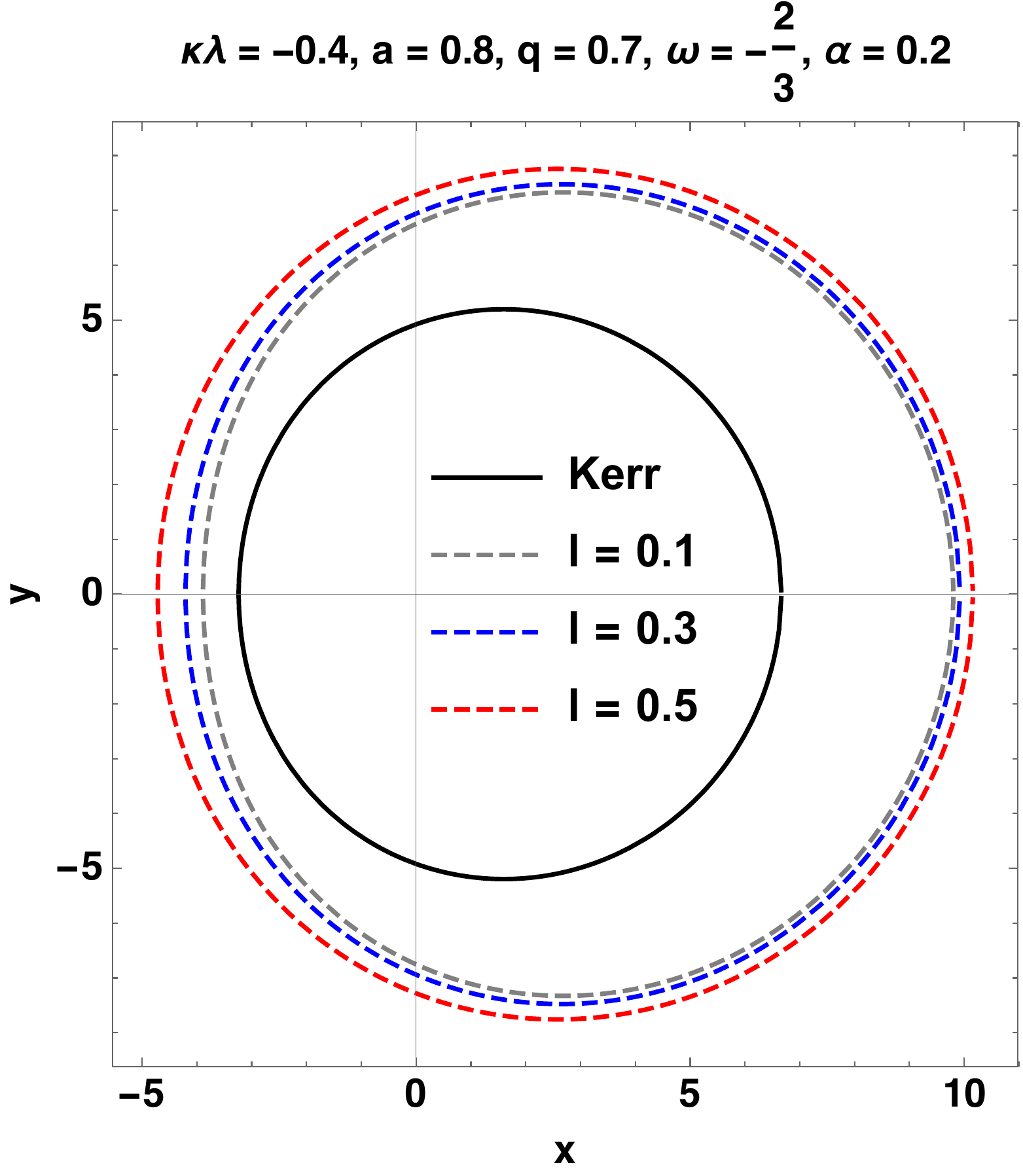}
		\hspace{2 cm}
		\includegraphics[width=0.355\linewidth]{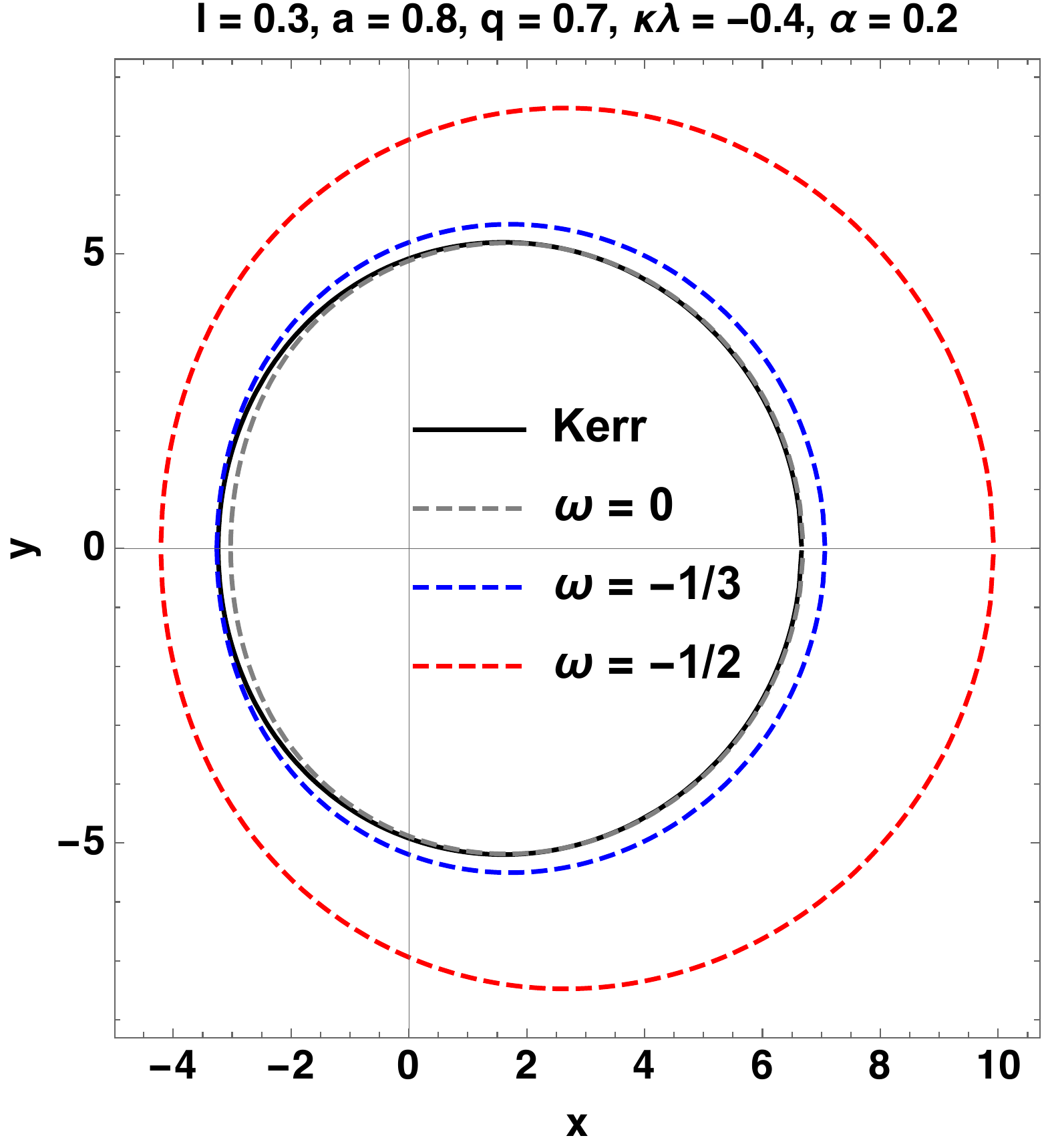}
		\includegraphics[width=0.36\linewidth]{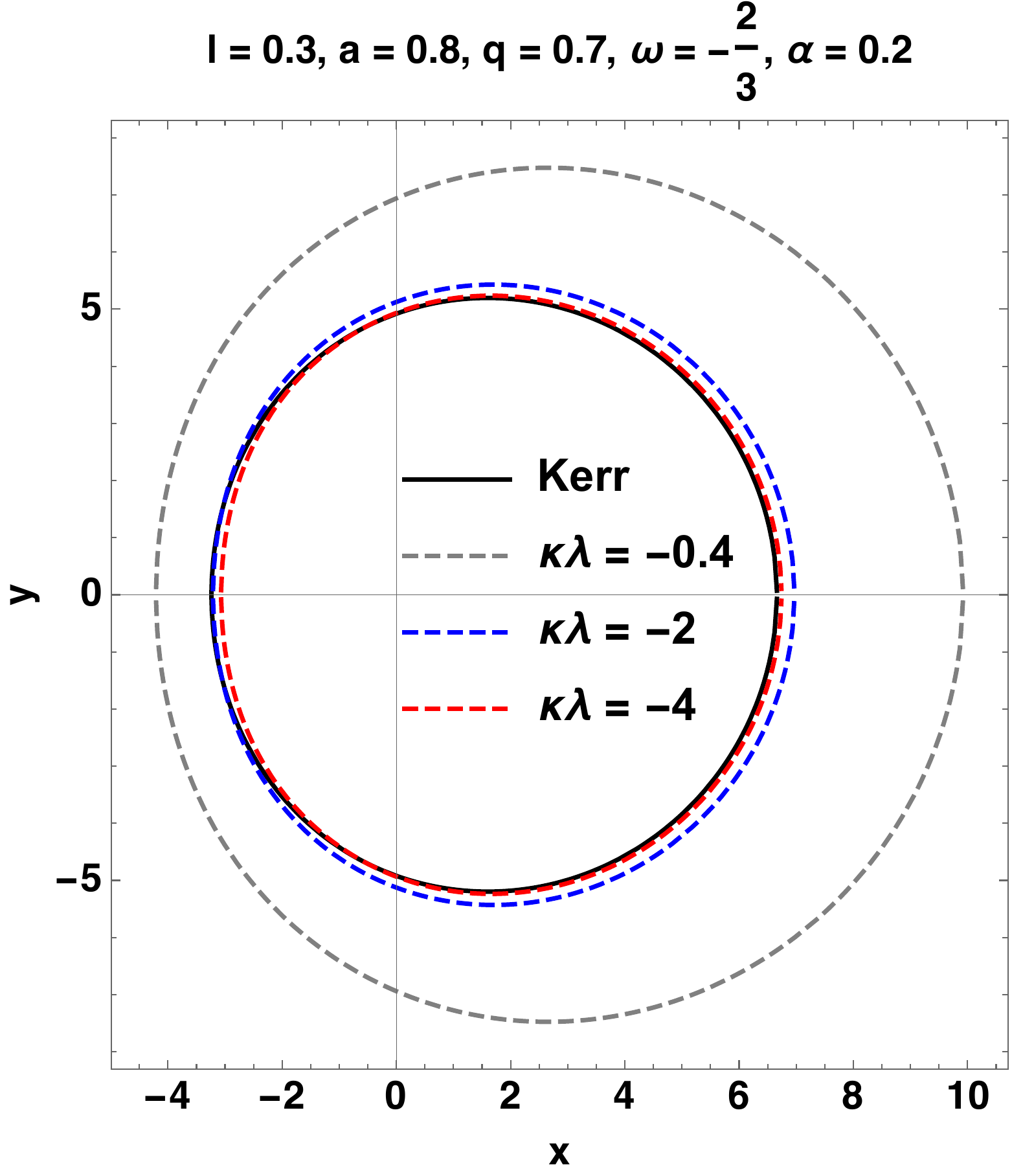}
		\hspace{1.8 cm}
		\includegraphics[width=0.379\linewidth]{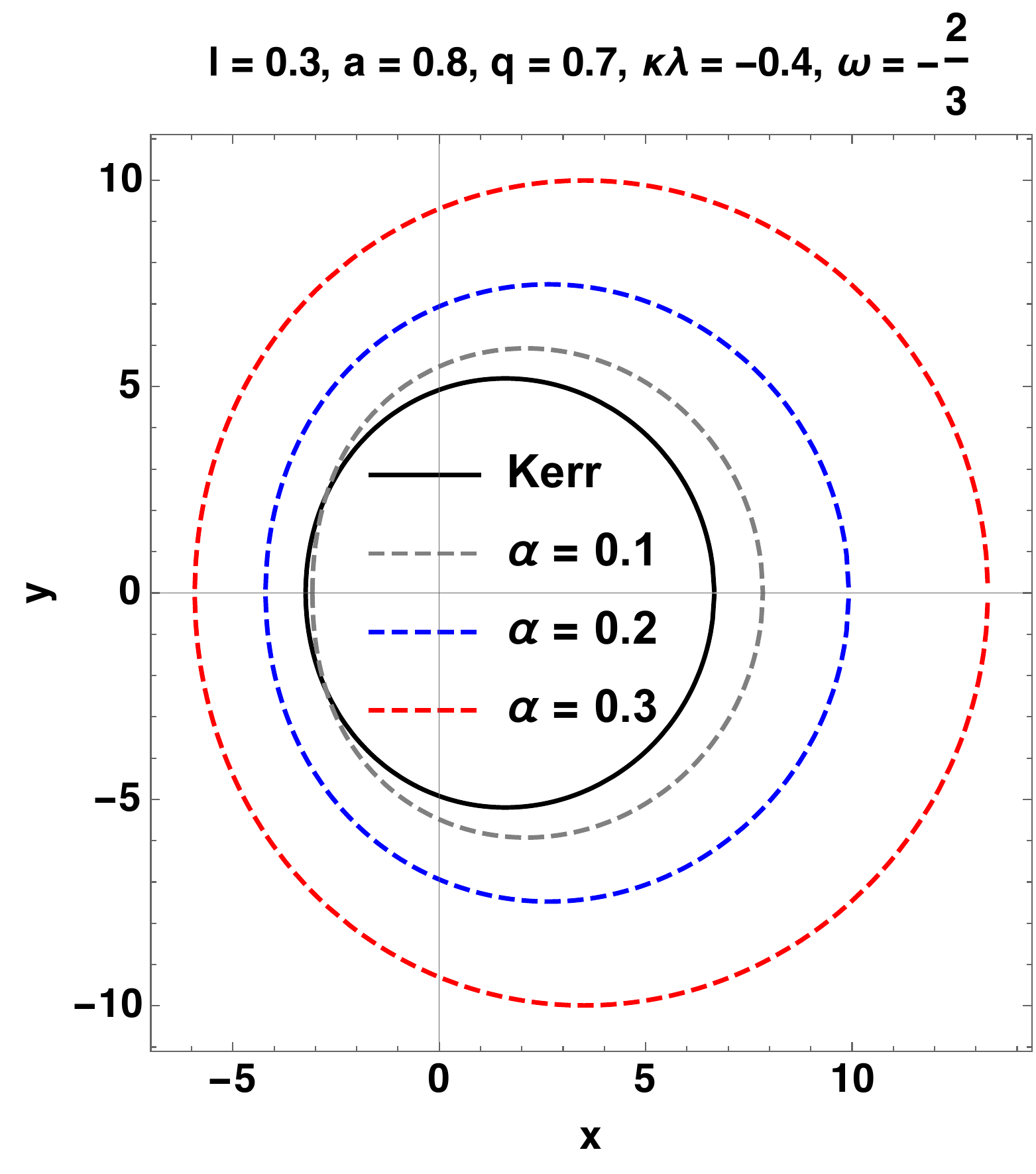}
		\includegraphics[width=0.39\linewidth]{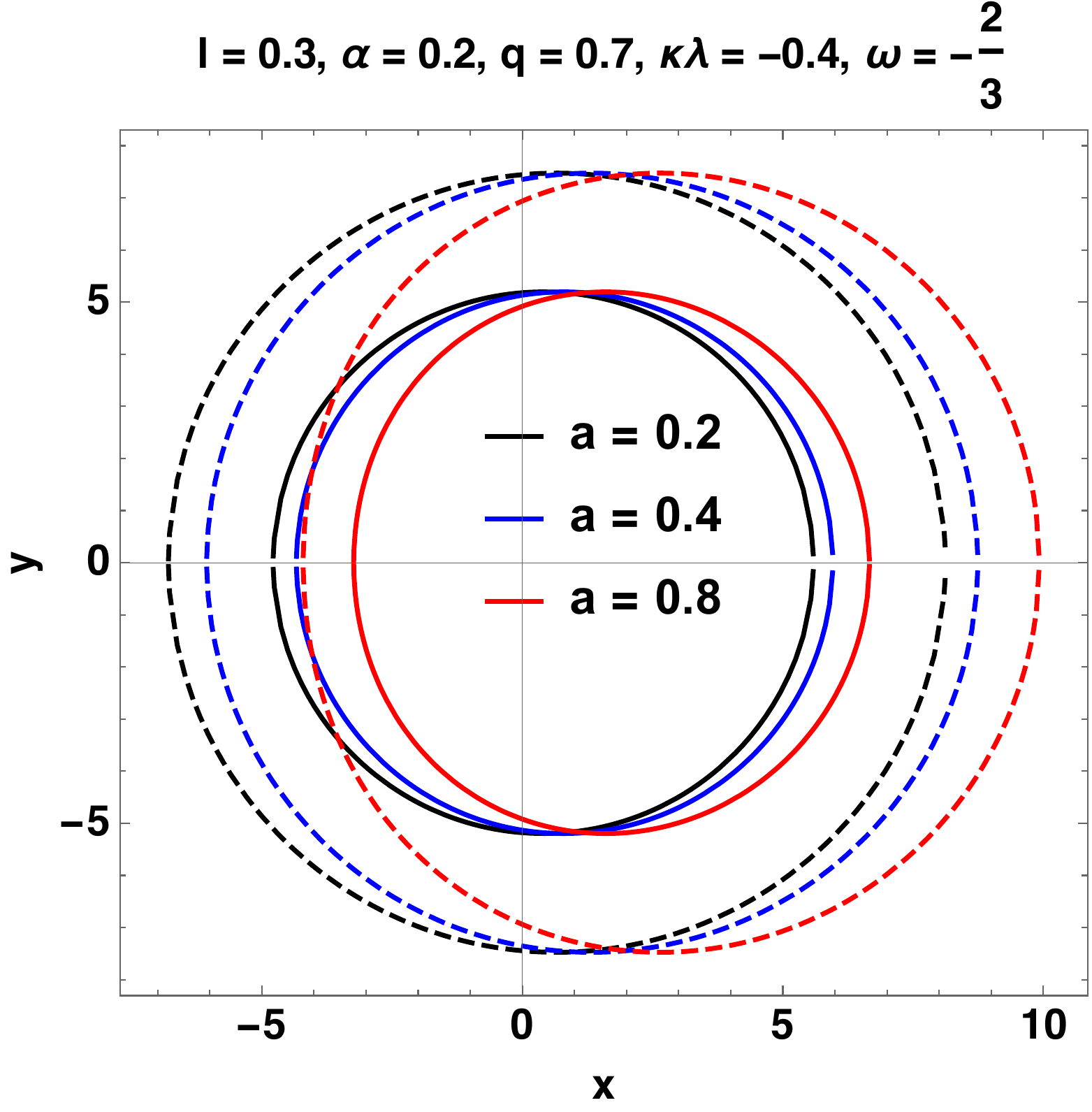}
		\hspace{1.8 cm}
		\includegraphics[width=0.36\linewidth]{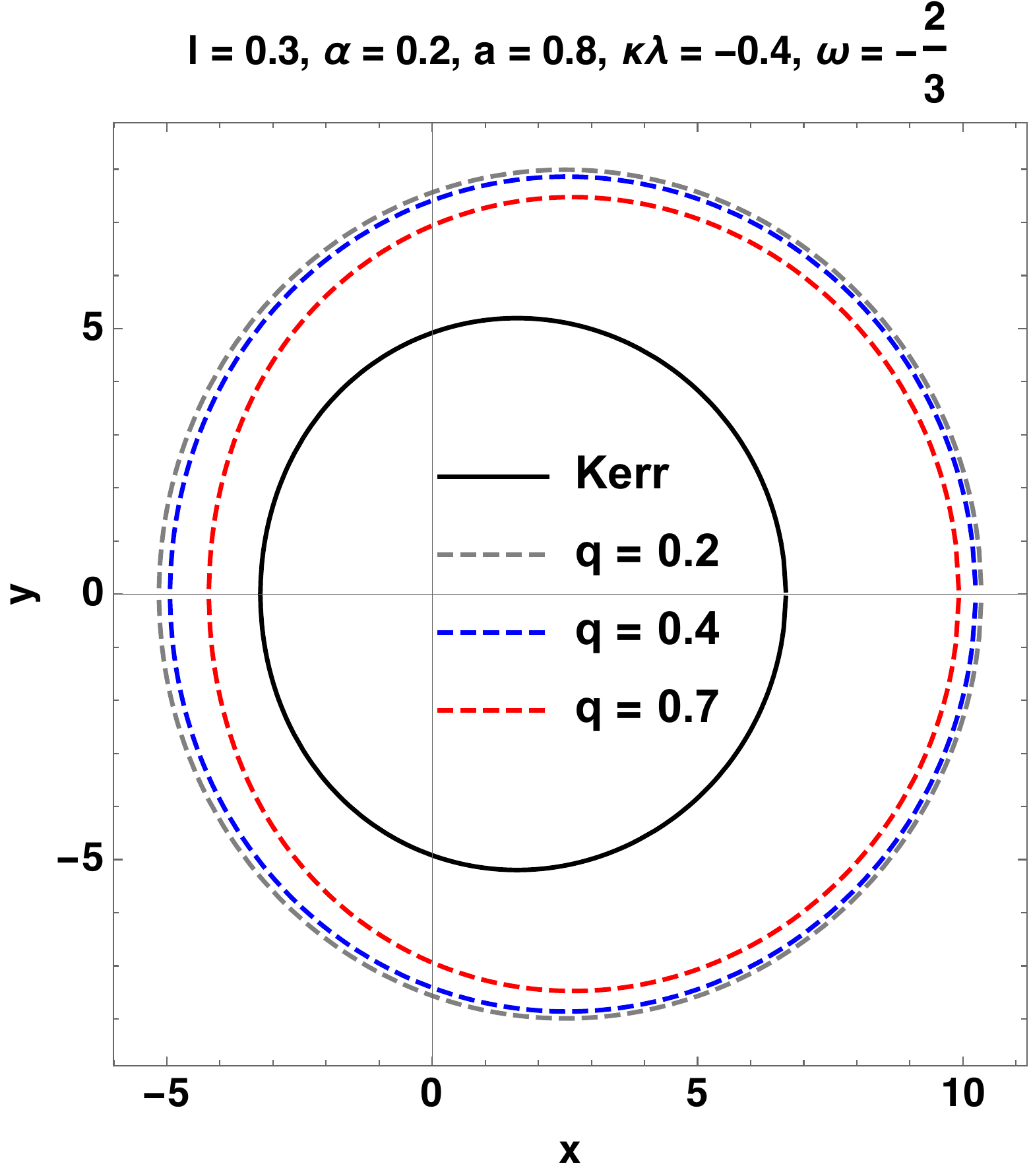}
	\end{center}
	\caption{The shape of the shadow of the CRNK black hole in the Rastall gravity observed at infinity is compared with the shadow of the Kerr black hole. \label{shw}}
\end{figure*}

Using the expressions for $x$ and $y$ one may easily get the shape of the shadow of the CRNK black hole in the Rastall gravity. We present the shadows of CRNK black hole in the Rastall theory of gravity in Fig.~\ref{shw}. 
In the top left plot of Fig.~\ref{shw} the change in the shape and size of the shadow around CRNK black hole in the Rastall gravity for the different values of the NUT charge $l$ is shown. In the same plot one can see that increase of the parameter $l$ increases the average radius of the shadow. Here one can also see that for the same spin and mass parameters the shadow cast by the CRNK black hole in the Rastall gravity is considerably larger than that of the Kerr black hole, corresponding to the fixed values of the rest of the spacetime parameters of the CRNK black hole in the Rastall gravity. In the top right plot of Fig.~\ref{shw} one can see the increase in the size of the shadow of the CRNK black hole in the Rastall gravity for the decreasing values of the parameter $\omega$. It is also noticeable that for the values of this parameter $\omega$ close to zero, the size of the shadow formed around the CRNK black hole in the Rastall gravity is almost the same as seen in the case of the Kerr black hole. In the left plot of the second row in the same figure the shape and size of the shadow of the CRNK black hole in the Rastall gravity for the various values of the parameter $\kappa\lambda$ are presented. From the same plot it can be easily noticed that for smaller values of this parameter $\kappa\lambda$, the size of the shadow of the CRNK black hole in the Rastall gravity coincides with the one for the Kerr black hole. This and previous plots show that for smaller values of the parameter $\kappa\lambda$, for the values of the parameter $\omega$ close to zero, it becomes difficult to distinguish the apparent shadow of the CRNK black hole in the Rastall gravity from the shadow of the the Kerr black hole. One can see strong dependence of the size of the shadow of the CRNK black hole in the Rastall gravity on the parameter $\alpha$, as it increases considerably with the increase of this parameter $\alpha$ and is shown in the second plot of the second row of the Fig.~\ref{shw}. It is demonstrated that for the values of the parameter $\alpha$ close to zero the size of the shadow of the CRNK black hole in the Rastall gravity becomes identical to the shadow of the Kerr black hole. In the left plot of the last row of Fig.~\ref{shw} the shadow of the CRNK black hole in the Rastall gravity is compared with the shadow of the Kerr black hole for the different values of the spin parameter $a$. It is mentioned here that in the Fig.~\ref{shw} solid lines correspond to the Kerr case and the dashed lines correspond to the CRNK black hole case in the Rastall gravity. Even though the values of the spin parameter $a$ are shown only for the Kerr case, the dashed lines with the same color represent the shadow with the same spin parameter $a$ for the CRNK black hole in the Rastall gravity. One can easily see that increase of the spin parameter $a$ does not change the size of the shadow much and only shifts the center of the circle of the shadow to the right in both the Kerr and CRNK spacetimes. In the last plot of the Fig.~\ref{shw}, it is shown how the electromagnetic charge $q$ affects the shadow around the CRNK black hole in the Rastall gravity. One can see that increase in the parameter $q$ reduces the size of the shadow. From the discussion presented here we conclude that when the gravitational field around a black hole becomes stronger the size of the shadow increases and when it becomes weaker the size of the shadow decreases.

\begin{figure*}[h!]
	\begin{center}
		\includegraphics[width=0.39\linewidth]{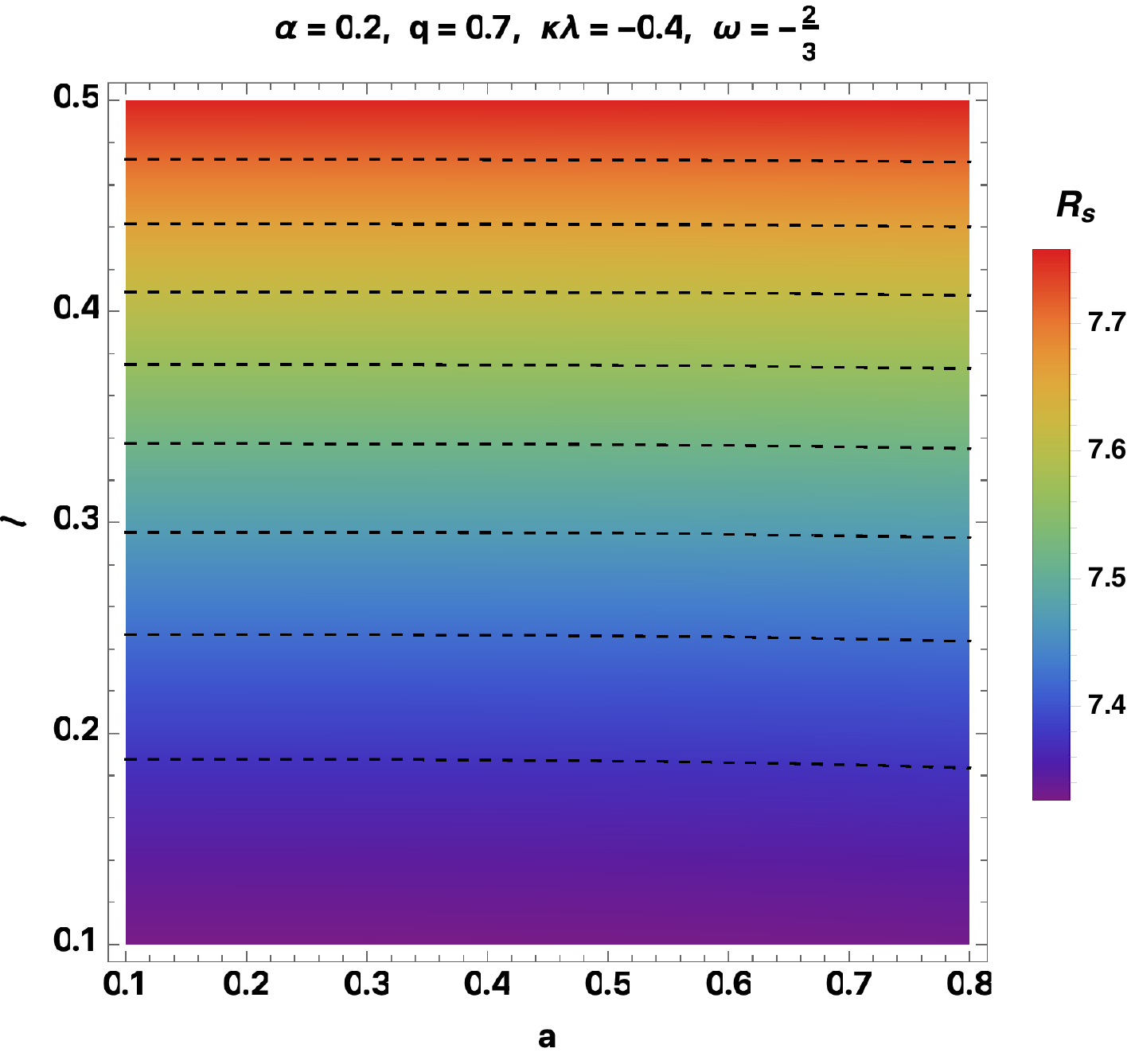}
		\hspace{2 cm}
		\includegraphics[width=0.39\linewidth]{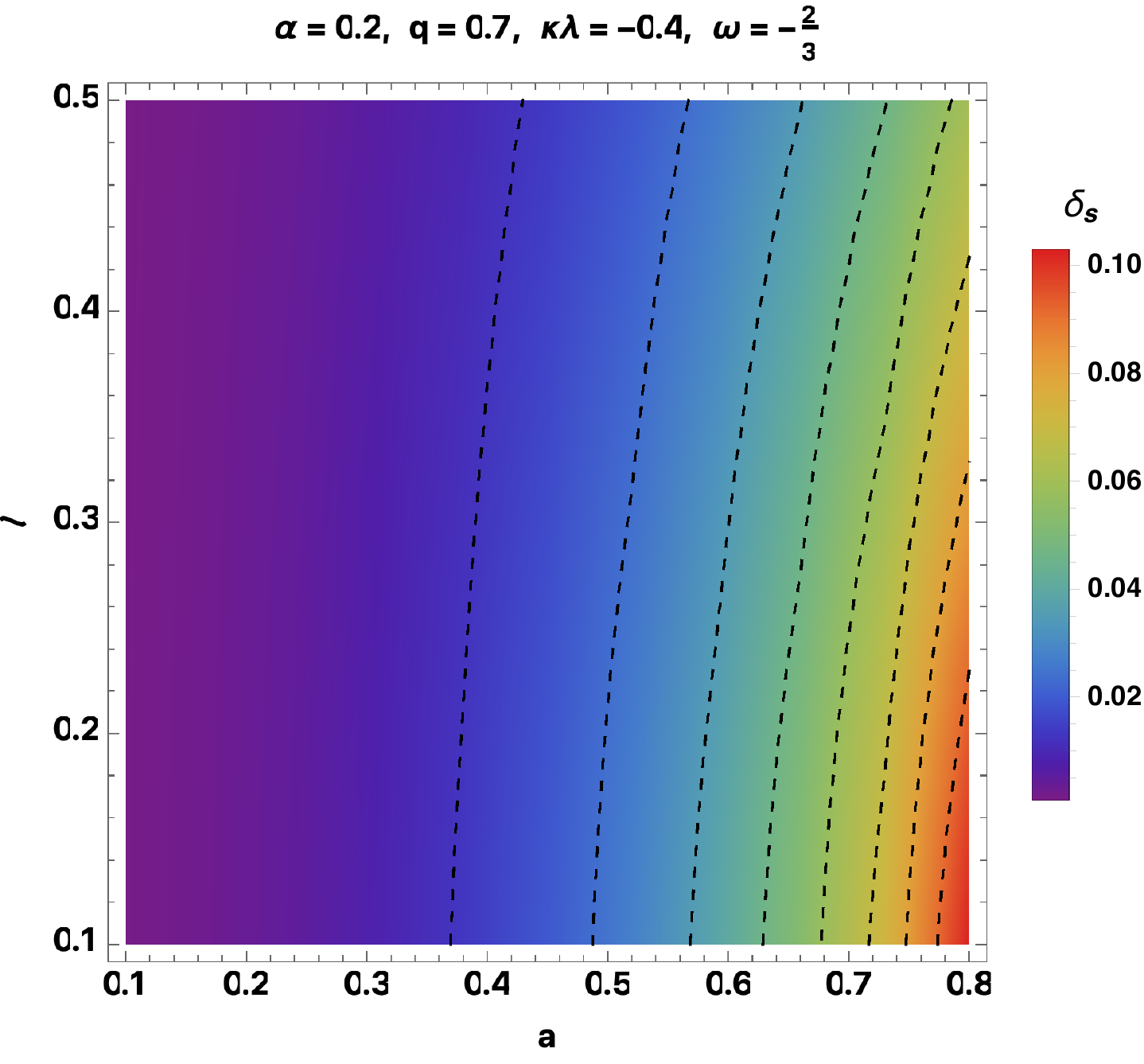}\vspace{1 cm}
		\includegraphics[width=0.39\linewidth]{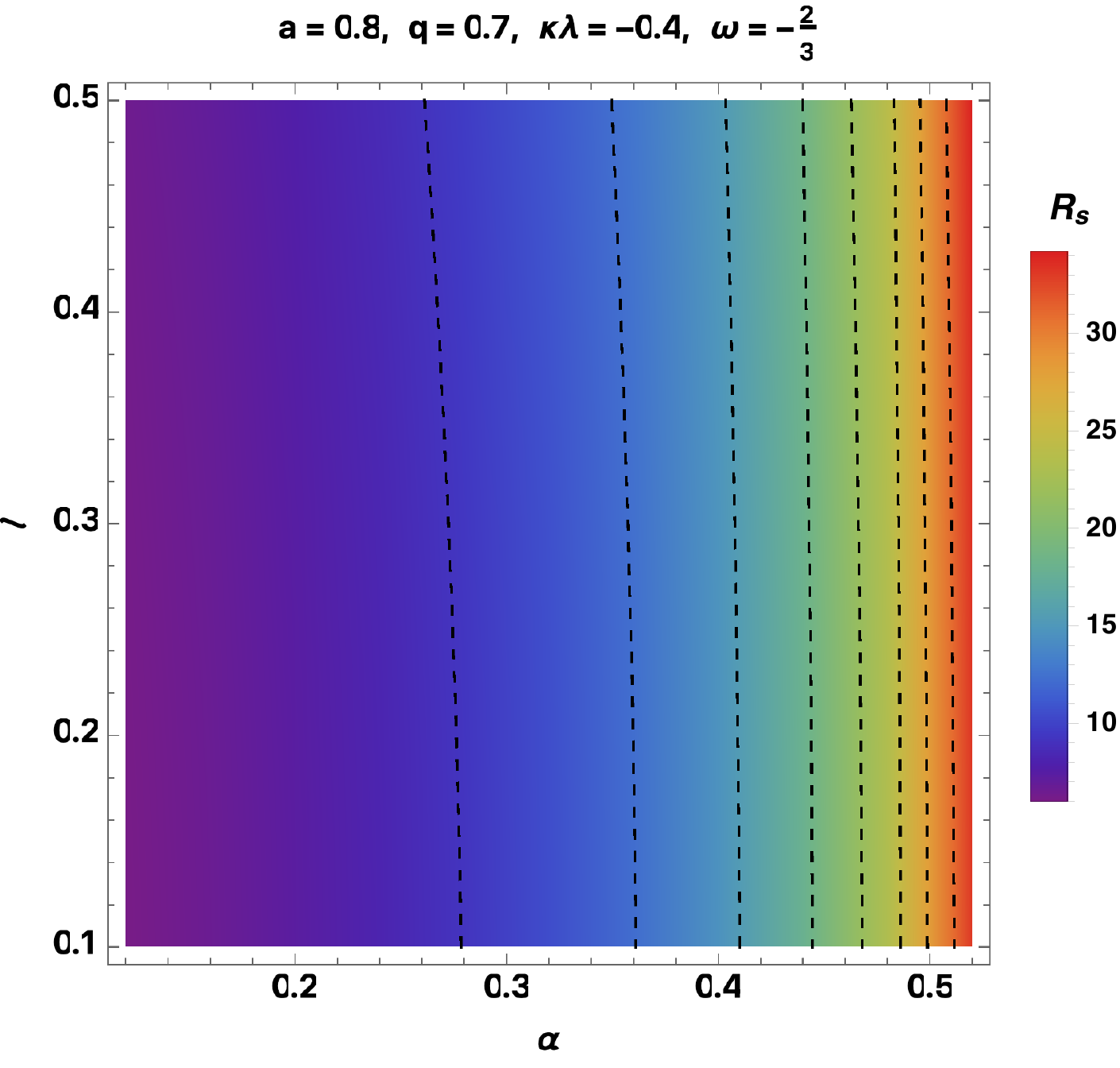}
		\hspace{1.8 cm}
		\includegraphics[width=0.39\linewidth]{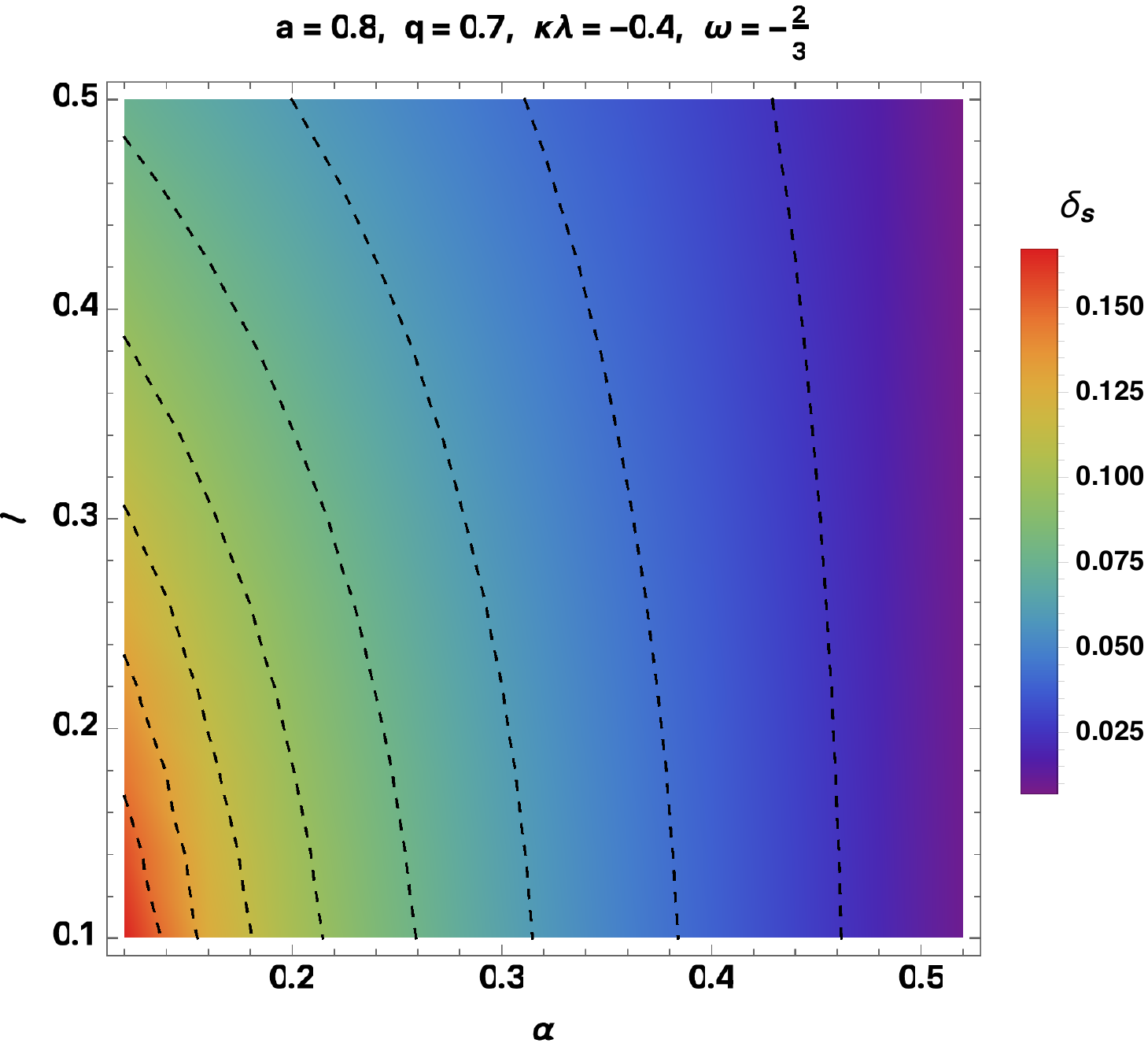}\vspace{1 cm}
		\includegraphics[width=0.39\linewidth]{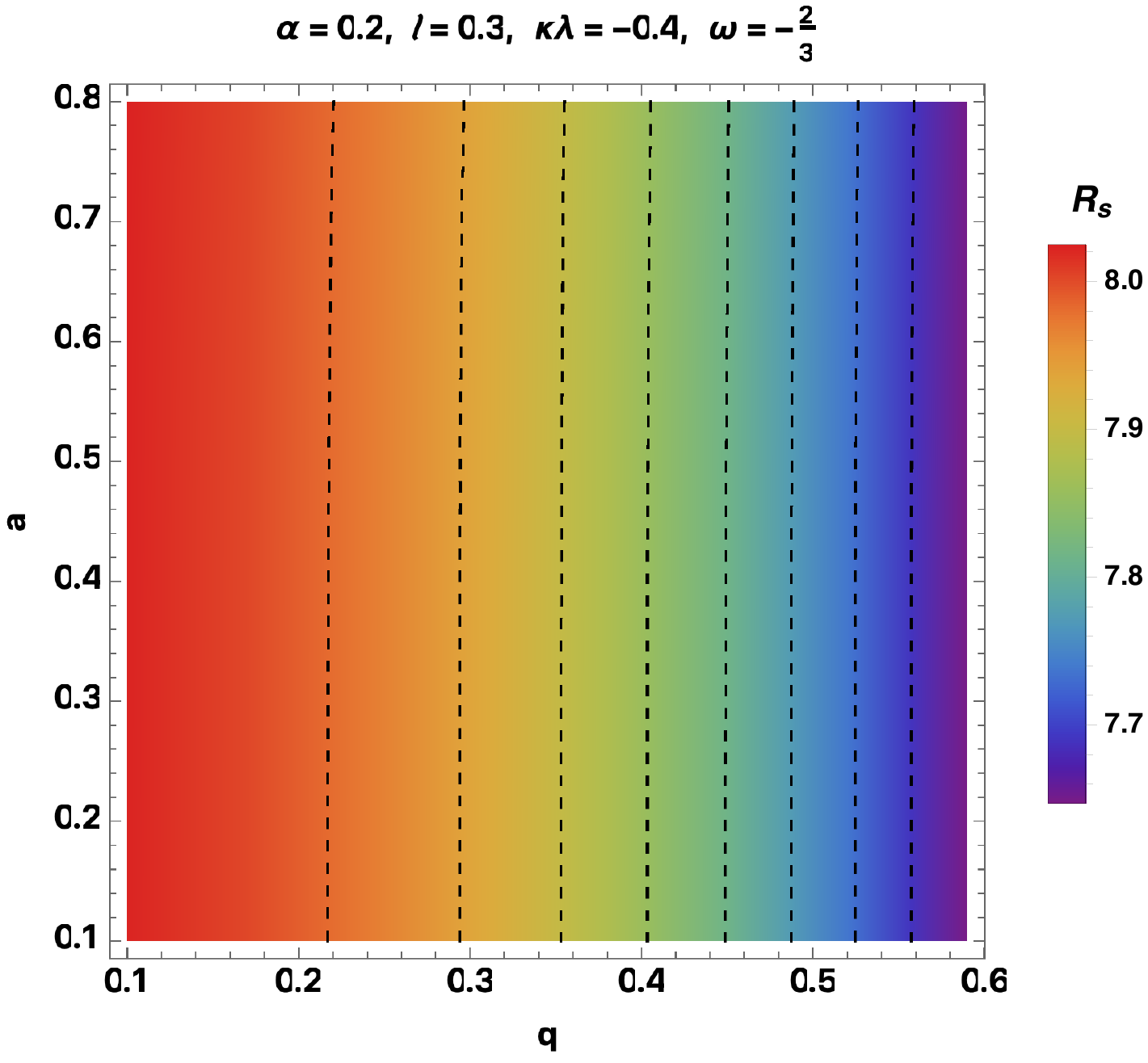}
		\hspace{1.8 cm}
		\includegraphics[width=0.39\linewidth]{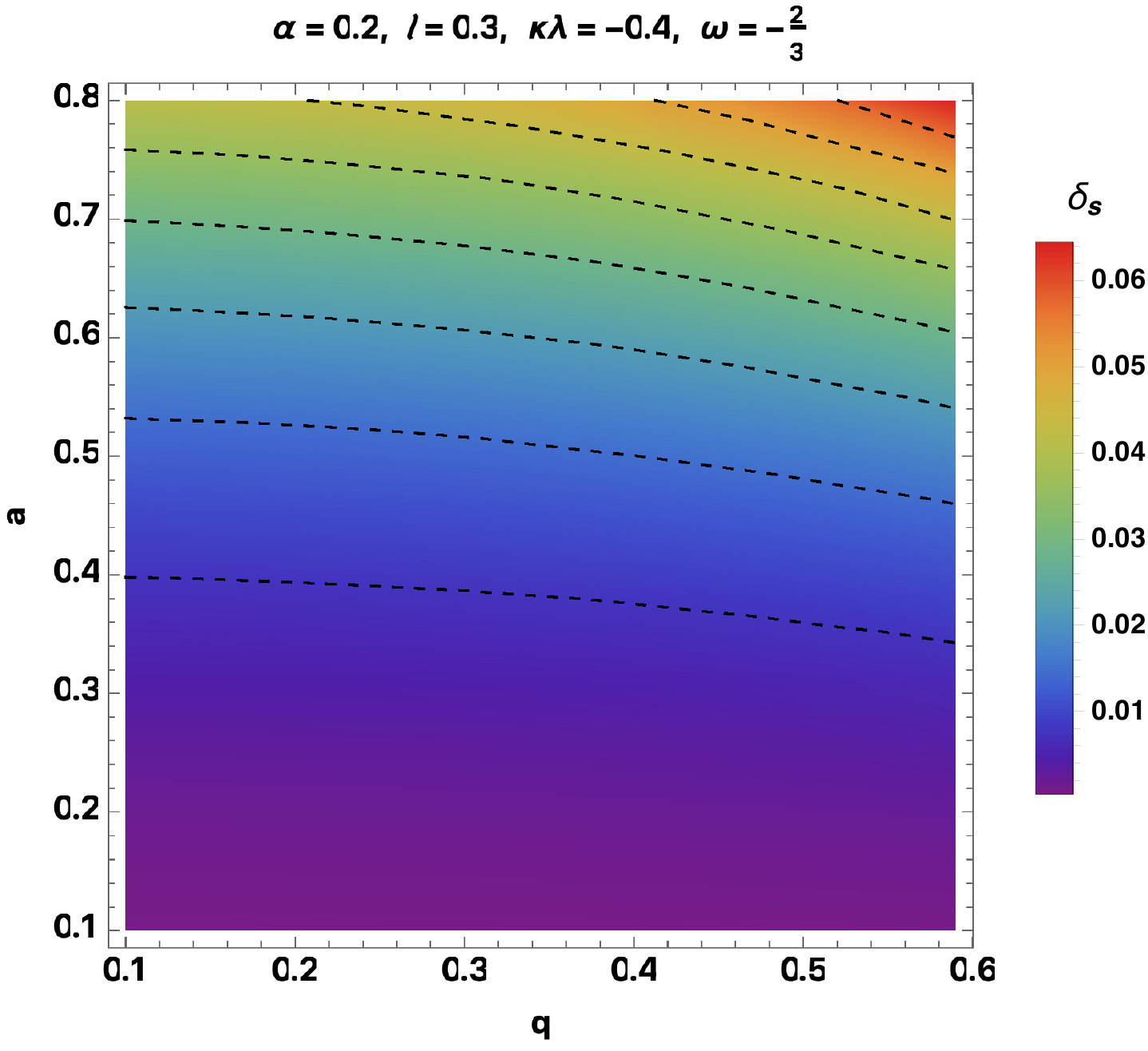}
	\end{center}
	\caption{The dependence of $R_s$ and $\delta_s$ on the spacetime parameters. \label{Rsds}}
\end{figure*}

\begin{figure*}[h!]
	\begin{center}
		\includegraphics[width=0.39\linewidth]{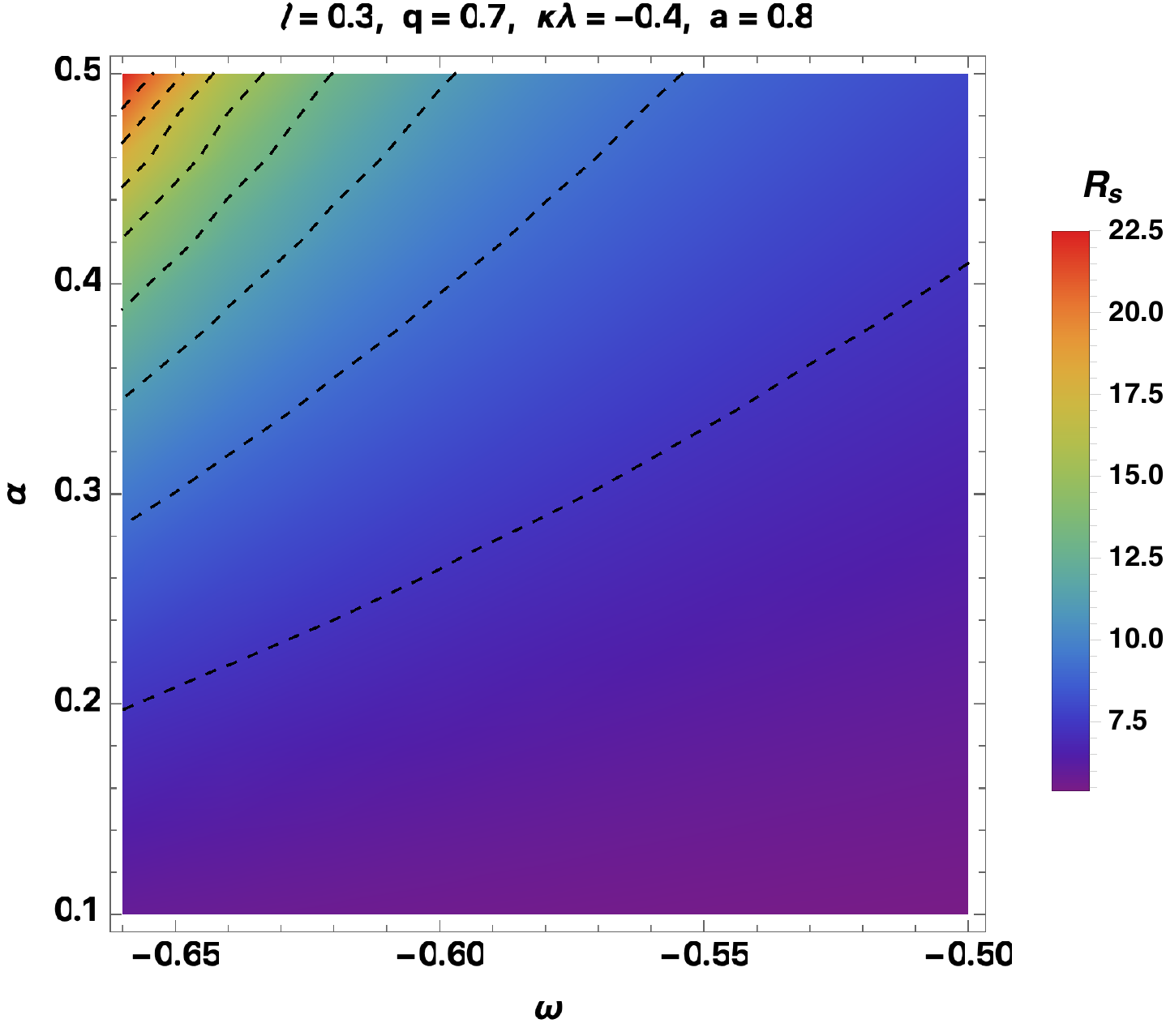}
		\hspace{2 cm}
		\includegraphics[width=0.39\linewidth]{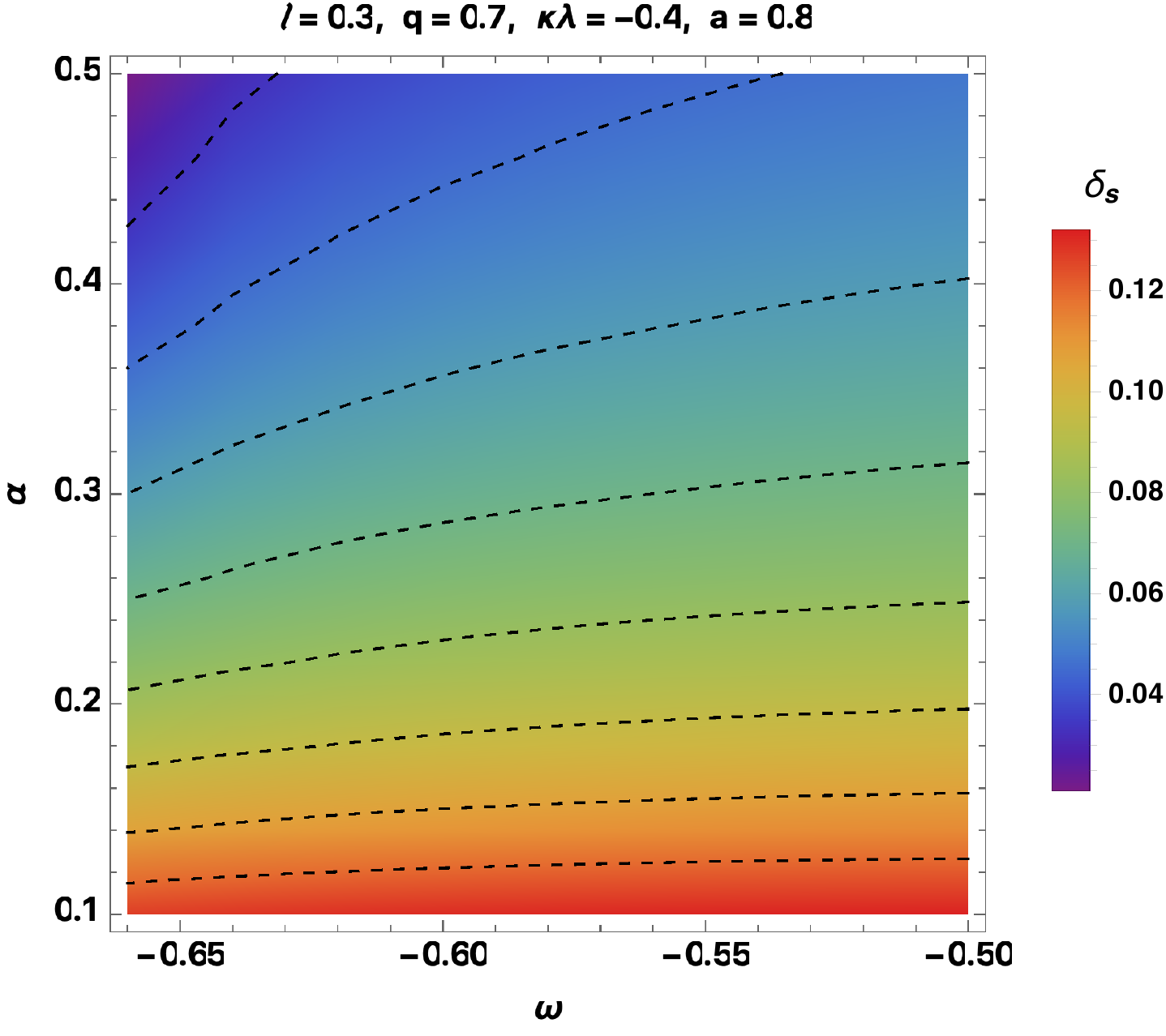}\vspace{1 cm}
		\includegraphics[width=0.4\linewidth]{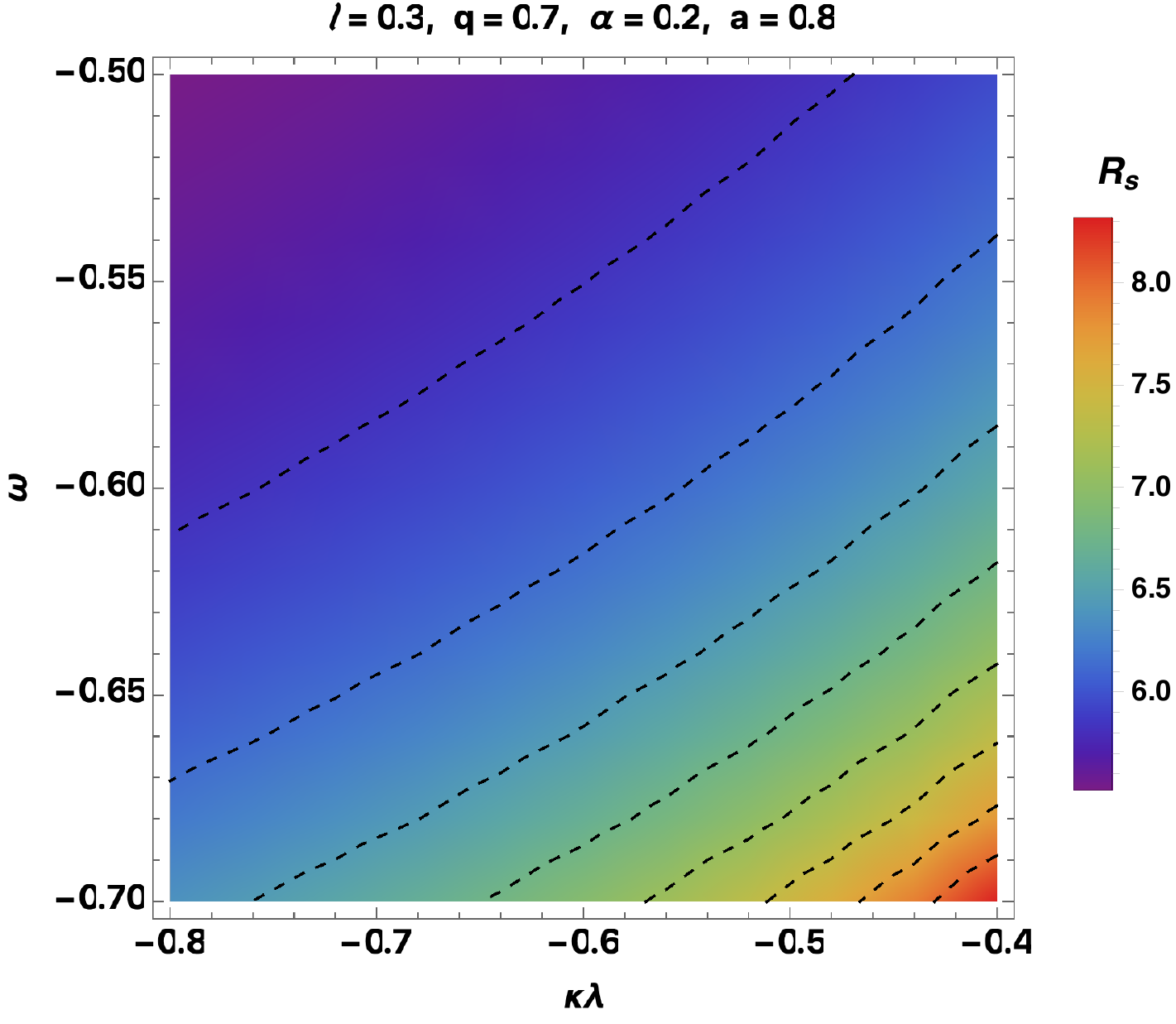}
		\hspace{1.8 cm}
		\includegraphics[width=0.4\linewidth]{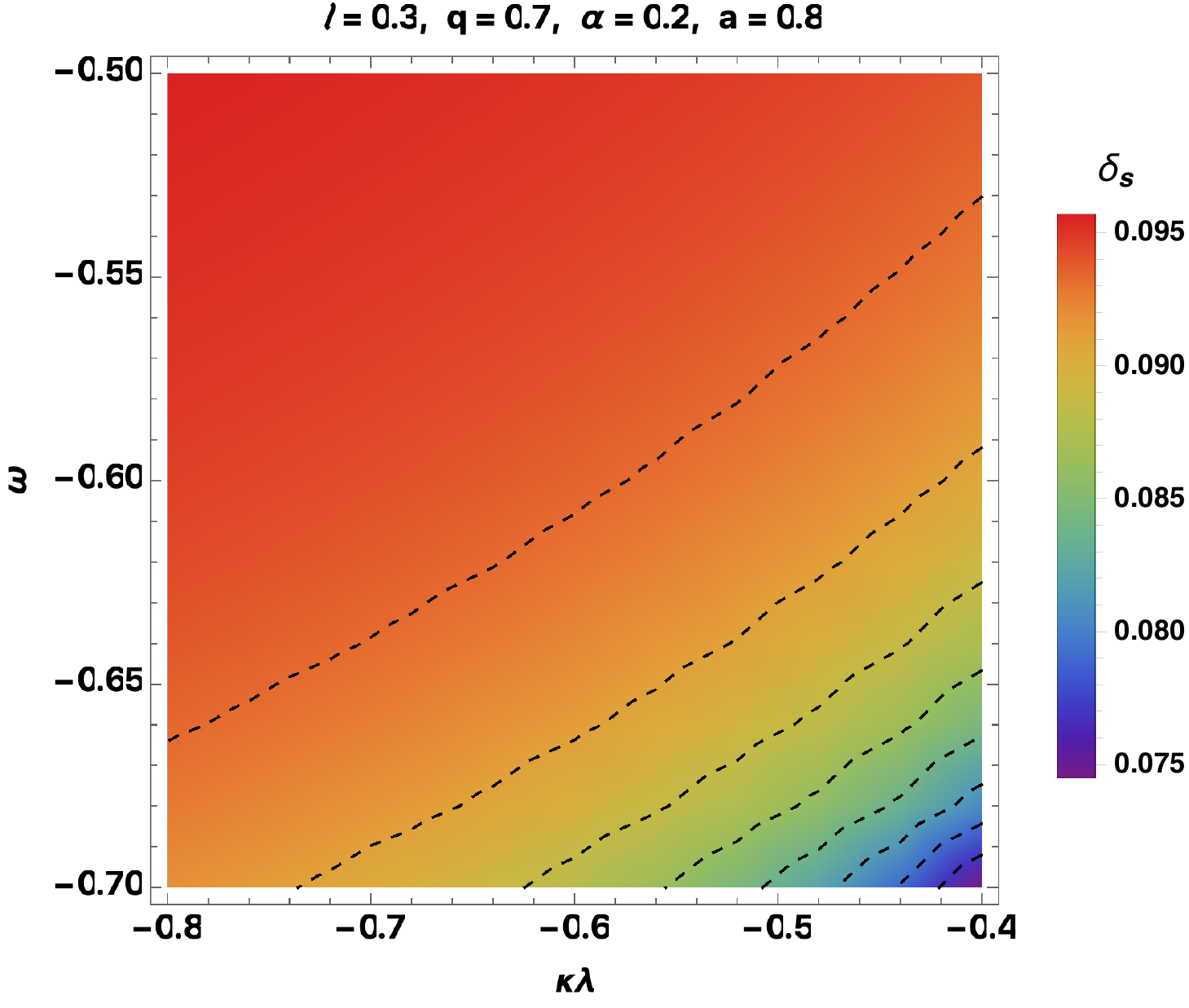}\vspace{1 cm}
		\includegraphics[width=0.39\linewidth]{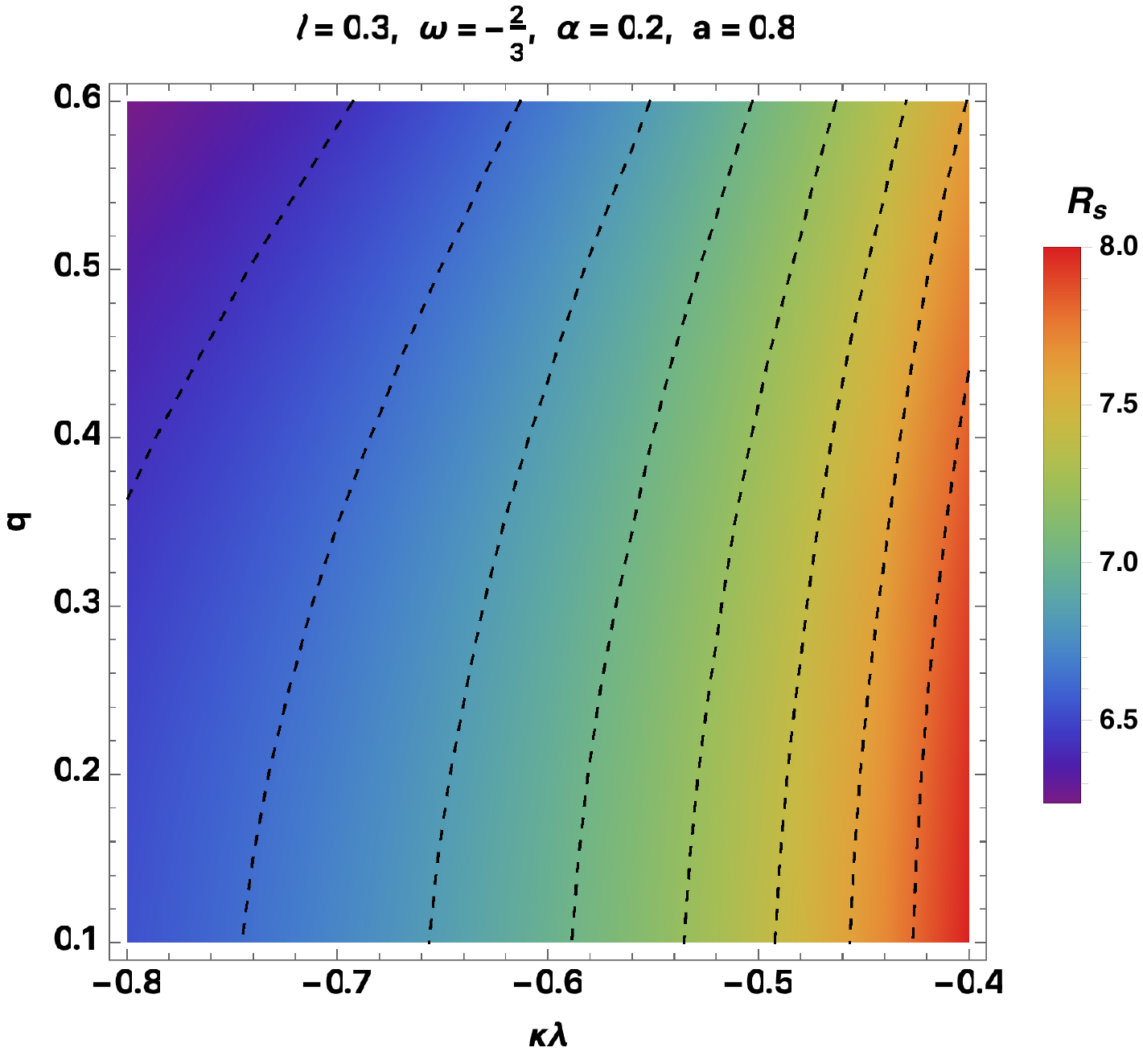}
		\hspace{1.8 cm}
		\includegraphics[width=0.39\linewidth]{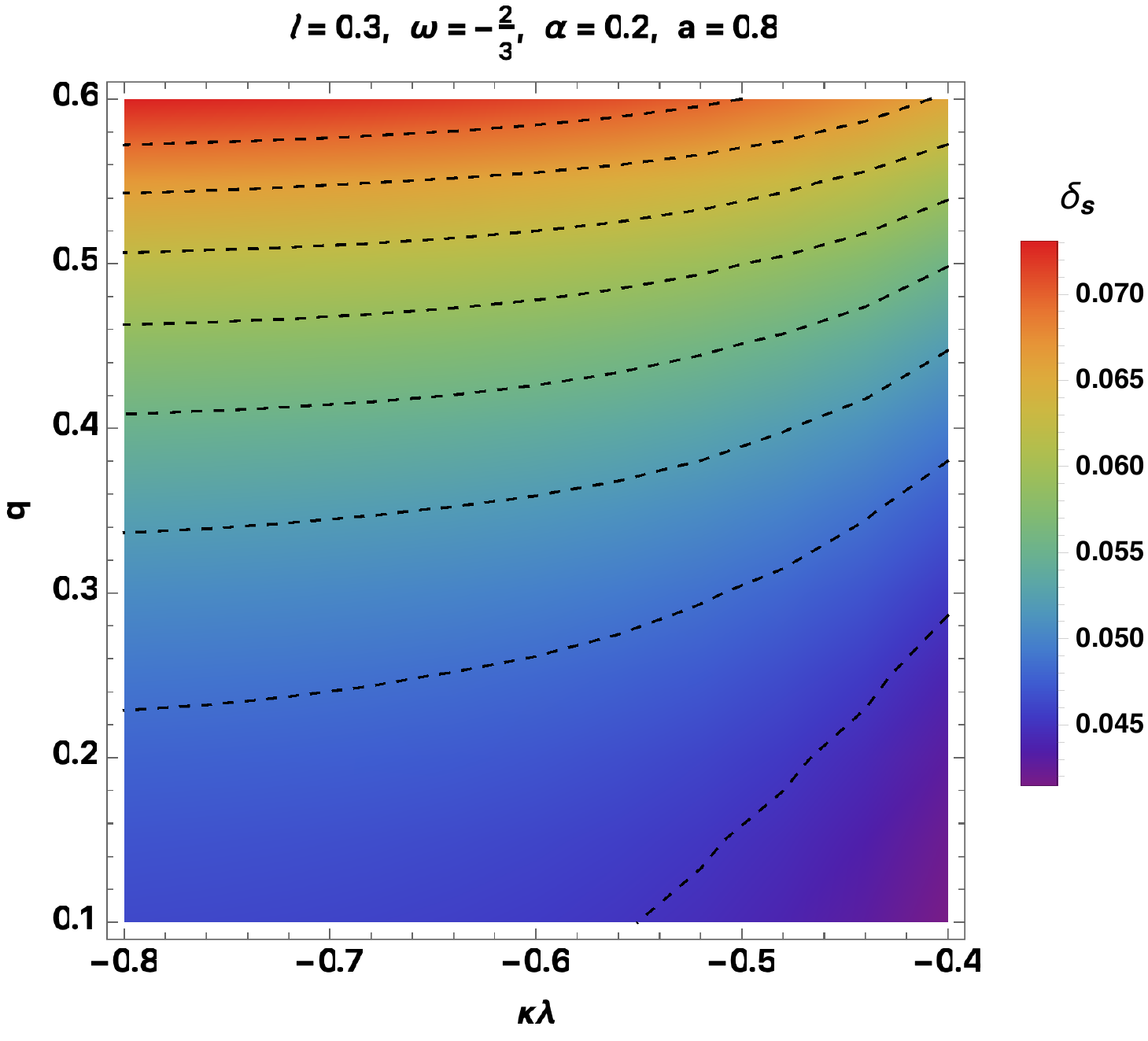}
	\end{center}
	\caption{The dependence of $R_s$ and $\delta_s$ on the spacetime parameters as in the previous figure. \label{Rsds2}}
\end{figure*}

One can obtain the observable $R_s$ using the equation
\begin{eqnarray}
R_s=\frac{(x_C-x_A)^2+y_A^2}{2 (x_C-x_A) }
\end{eqnarray}
and the other observable $\delta_s$ can be calculated using the relation
\begin{eqnarray}
\delta_s=\frac{D_{cs}}{R_s}.
\end{eqnarray}
The observables $R_s$ and $\delta_s$ are presented in Figs.~\ref{Rsds} and~\ref{Rsds2} (plots on the left for $R_s$ and plots on the right for $\delta_s$) for the different values of the spacetime parameters of the CRNK black hole in the Rastall gravity. In the top left plot of the Fig.~\ref{Rsds} it is shown that the average shadow radius $R_s$ increases with the increase of the NUT parameter $l$. There is a negligible decrease with the increase of the spin parameter $a$ in $R_s$. In the right plot of the first row of the Fig.~\ref{Rsds} one can see deviation of the shape of the shadow around CRNK black hole in the Rastall gravity from a circle. This deviation becomes more noticeable for smaller values of the NUT charge $l$ and for faster rotation of the CRNK black hole in the Rastall gravity. From the left plot of the second row of the Fig.~\ref{Rsds} one can see that the effect of the parameter $\alpha$ on the average radius of the shadow of the CRNK black hole in the Rastall gravity is much stronger as compared to the effects of the NUT parameter $l$. This is due to the fact that $R_s$ goes up faster with the increase of $\alpha$ than the increase in the NUT parameter $l$. However, with the increase of parameters $\alpha$ and $l$ the deviation of the shape of the shadow from a circle becomes smaller as shown in the right plot of the same row in the same figure. In the left plot of the third row of the Fig.~\ref{Rsds} we have shown how the average radius of the black hole shadow $R_s$ changes with the change of the spin parameter $a$ and the charge parameter $q$. One can easily see that the effect of the latter is stronger as compared to the effect of the spin of the CRNK black hole in the Rastall gravity. In the right plot of the same row in the same figure it is clearly seen that for bigger values of the parameters $q$ and $a$ the deviation of the shape of the shadow from a circle becomes more visible.

In the first plot of Fig.~\ref{Rsds2} we have demonstrated how the observable $R_s$ changes due to the change in parameters $\alpha$ and $\omega$. One can see that smaller values of $\omega$ and bigger values of $\alpha$ produce bigger average radius of the shadow around the CRNK black hole in the Rastall gravity. In the right plot of the first row of the Fig.~\ref{Rsds2} it is shown that for smaller values of the parameter $\alpha$, the effect of the parameter $\omega$ on the observable $\delta_s$ is negligible. Moreover, we have shown that the distortion in the black hole shadow becomes more apparent only for the bigger values of the parameter $\alpha$. In the left plot of the second row of the Fig.~\ref{Rsds2} it is shown that for smaller values of $\omega$ and bigger values of $\kappa\lambda$ the observable $R_s$ takes bigger values. In the right plot of the same row in the Fig.~\ref{Rsds2} it is clearly seen that the second observable $\delta_s$ becomes smaller in the corresponding range of values of the spacetime parameters for the CRNK black hole in the Rastall gravity. In the last row of the Fig.~\ref{Rsds2} the change of the observables $R_s$ and $\delta_s$ in the $\kappa\lambda - q$ plane is demonstrated. It can be seen that for smaller values of $q$ and bigger values of $\kappa\lambda$ the observable $R_s$ becomes bigger while the observable $\delta_s$ becomes smaller i.e. the average radius of the shadow of the CRNK black hoe in the Rastall gravity increases and the black hole shadow almost assumes the shape of a perfect circle.

\section{Conclusion}

In this work we have investigated photon motion, the GL and the black hole shadow in the spacetime of the CRNK black hole in the Rastall theory of gravity. For the photon motion we have used the Hamilton-Jacobi equation of motion for massless particles. From the relation between the radius of closest approach of photons to the central black hole and the impact parameter of photons we have found  that this relation becomes approximately  linear  for the bigger values of these quantities. We have shown that for the fixed value of the impact parameter, with the increase of the values of the parameters $l$, $\alpha$, and $\kappa\lambda$ the radius of the closest approach of photons to the central object goes down. We have observed the opposite effect on the radius of the closest approach with the increase in the values of the rest of the spacetime parameters. We have evaluated the bending angle of photons due to the gravitational field of the central black hole and  have shown that the effect of the spacetime parameters $l$, $\omega$, and $\alpha$ is much stronger as compared to the effect of the parameters $a$, $\kappa\lambda$ and $q$. We have demonstrated that increase in $l$ and $\alpha$ increases the bending angle while increase in $\omega$ reduces it. In spite of the fact that the effects of the parameters $a$, $q$, and $\kappa\lambda$ are negligible, we have demonstrated that increase of the first two parameters reduces the bending angle and increase of the third parameter increases it. We have also studied the influence of the spacetime parameters on the photon sphere formed around the CRNK black hole in the Rastall gravity. It is shown that i) increase of the parameter $\alpha$ makes the radius of photon sphere greater and this increase in the radius of the photon sphere becomes faster with the increase of the NUT parameter $l$. On the contrary the radius of the photon sphere decreases with the decrease of the Rastall parameter $\kappa\lambda$; ii) the radius of the photon sphere goes up exponentially when the parameter $\omega$ approaches the value $\sim-0.8$; iii) the effect of the Rastall parameter $\kappa\lambda$ on the radius of the photon sphere strongly depends on the parameters $\omega$ and $\alpha$. Further we have observed that for bigger values of $\alpha$ and smaller values of $\omega$ the effect becomes stronger; iv) the effect of the parameters $a$ and $q$ on the radius of the photon sphere is similar and increase in the values of both the parameters reduces the radius of the photon sphere; v) the radius of the photon sphere becomes bigger for bigger values of the parameters $l$ and $\kappa\lambda$.\\

Then we have investigated the shadow formed around the CRNK black hole in the Rastall gravity. We have constructed the shape of the shadow in comparison with the Kerr black hole case. We have obtained the following results i) the size of the shadow slightly increases with the increase of the NUT parameter $l$ and it decreases with the increase of the electromagnetic charge parameter $q$; ii) decrease in the value of the spacetime parameter $\omega$ makes the size of the shadow bigger; iii) decrease in the value of the parameter $\kappa\lambda$ makes the size of the black hole shadow smaller. Further the difference in the sizes of the shadows for the values of the parameter $\kappa\lambda$ between -2 and -4 is negligible. The size of the shadow in the case $\kappa\lambda=-0.4$ is much bigger than these two values of the $\kappa\lambda$ i.e. -2 and -4; iv) the influence of the parameter $\alpha$ on the size of the shadow is much stronger as compared to the effect of the rest of the spacetime parameters. The increase of the  parameter $\alpha$ causes increase in the size of the shadow; v) the change of the spin parameter $a$ only changes the position of the black hole shadow. The change in the size of the shadow of the black hole is negligible for change in the parameter $a$. As the final step we have studied the dependence of the two observable quantities, $R_S$ 
and $\delta_s$ 
on the spacetime parameters. We have demonstrated that the effect of the spin parameter $a$ on the size of the shadow is indeed negligible but can change the distortion parameter $\delta_s$ considerably for smaller values of the NUT parameter $l$ and for the bigger values of the electromagnetic charge parameter $q$. We have seen that for bigger values of the parameter $l$ the observable $R_s$ becomes bigger but the second observable $\delta_s$ becomes smaller. Similar behavior of the two observable is observed for different values of the parameters $\alpha$ and $\kappa\lambda$. We have seen that the bigger values of $R_s$ corresponds to smaller values of the $\delta_s$ and this effect is more sensitive to the values of the parameters $\omega$ and $\kappa\lambda$ in comparison to the other spacetime parameters.

 {One could ask the question whether is it possible to get constraints on the parameters of a CRNK black hole in the Rastall gravity using the M87* black hole shadow observations by the EHT collaboration. The low precision of the M87* black hole shadow image observation indicates, if we assume that it is the rotating Kerr black hole, the black hole spin measurement uncertainty will be in the big range within $0.5 \leq \alpha_{*} \leq 0.94$, where $\alpha_{*}=J / M^2$ ($J$ is the total angular momentum of the black hole). It means that with the current accuracy of the shadow observations the CRNK black  hole in the Rastall gravity is not distinguishable from the Kerr black hole.}

It is worth noticing that the results presented in this work are consistent with each other i.e. the effect of all the spacetime parameters on the GL, shadow cast by the black hole and  two observables $R_s$ and $\delta_s$ are in agreement. In the next stage we would like to investigate the shadow cast by the CRNK black hole in the Rastall gravity by comparing the images of the shadow formed from the radiation of an accretion disc around the black hole, obtained by two different ray tracing codes namely, "GYOTO" \cite{Vincent2011wz} and ray tracing code by Ayzenberg et. al \cite{Ayzenberg2018jip}.

\section*{Acknowledgments}
This research is supported  by Grants F-FA-2021-432 and MRB-2021-527 of the Uzbekistan Ministry for Innovative Development and  by the Abdus Salam International Centre for Theoretical Physics under the Grant No.  OEA-NT-01.

\bibliographystyle{apsrev4-1}
\bibliography{gravreferences}

\end{document}